\documentclass[numberedappendix]{emulateapj}
\usepackage{apjfonts}

\newcommand{\etal}{et al.\,}

\slugcomment{Accepted for publication in the Astrophysical Journal}

\shorttitle{Properties of stellar clumps in $z \sim 2$ galaxies}
\shortauthors{F\"orster Schreiber \etal}

\begin{document}

\title{Constraints on the assembly and dynamics of galaxies.~
II.~Properties of kiloparsec-scale clumps in rest-frame optical
emission of $z \sim 2$ star-forming galaxies \,\altaffilmark{1}}

\author{N. M. F\"orster Schreiber\altaffilmark{2},
        A. E. Shapley\altaffilmark{3,4},
        R. Genzel\altaffilmark{2,5},
        N. Bouch\'e\altaffilmark{6,7,8,9},
        G. Cresci\altaffilmark{10},
        R. Davies\altaffilmark{2},
        D. K. Erb\altaffilmark{11},
        S. Genel\altaffilmark{2,12},
        D. Lutz\altaffilmark{2},
        S. Newman\altaffilmark{13},
        K. L. Shapiro\altaffilmark{13,14},
        C. C. Steidel\altaffilmark{15},
        A. Sternberg\altaffilmark{12},
        and L. J. Tacconi\altaffilmark{2}
}

\altaffiltext{1}{Based on observations made with the NASA/ESA
  {\em Hubble Space Telescope\/}, obtained at the Space Telescope
  Science Institute, which is operated by the Association of
  Universities for Research in Astronomy, Inc., under NASA
  contract NAS 5$-$26555. These observations are associated
  with program \# 10924.
  Also based on observations obtained at the Very Large Telescope
  of the European Southern Observatory, Paranal, Chile
 (ESO Programme IDs 073.B-9018, 074.A-9011, 075.A-0466, 076.A-0527,
  077.A-0576, 078.A-0600, 079.A-0341, 080.A-0330, and 080.A-0339).}
\altaffiltext{2}{Max-Planck-Institut f\"ur extraterrestrische Physik,
                 Giessenbachstrasse, D-85748 Garching, Germany}
\altaffiltext{3}{Department of Physics and Astronomy,
                 University of California, Los Angeles, CA 90095-1547}
\altaffiltext{4}{Packard Fellow}
\altaffiltext{5}{Department of Physics, Le Conte Hall, University of California,
                 Berkeley, CA 94720}
\altaffiltext{6}{Department of Physics, University of California Santa Barbara,
                 Santa Barbara, California, CA 93106-9530}
\altaffiltext{7}{CNRS,
                 Institut de Recherche en Astrophysique et Plan\'etologie
                 de Toulouse, 14 Avenue E. Belin, F-31400 Toulouse, France}
\altaffiltext{8}{Universit\'e de Toulouse; UPS-OMP; IRAP;
                 F-31400 Toulouse, France}
\altaffiltext{9}{Marie Curie Fellow}
\altaffiltext{10}{INAF-Osservatorio Astrofisico di Arcetri,
                 Largo E. Fermi 5, I-50125 Firenze, Italy}
\altaffiltext{11}{Department of Physics,
                  University of Wisconsin Milwaukee,
                  Milwaukee, Wisconsin, WI 53211}
\altaffiltext{12}{Sackler School of Physics \& Astronomy,
                 Tel Aviv University, Tel Aviv 69978, Israel}
\altaffiltext{13}{Department of Astronomy, Campbell Hall,
                  University of California,
                  Berkeley, CA 94720}
\altaffiltext{14}{Aerospace Research Laboratories,
                  Northrop Grumman Aerospace Systems,
                  Redondo Beach, CA 90278}
\altaffiltext{15}{California Institute of Technology, MS 105-24,
                  Pasadena, CA 91125}

\begin{abstract}

We study the properties of luminous stellar ``clumps'' identified in
deep, high-resolution {\em Hubble Space Telescope} NIC2/F160W imaging
at 1.6\,\micron\ of six $z \sim 2$ star-forming galaxies with existing
near-infrared integral field spectroscopy from SINFONI at the Very Large
Telescope.  Individual clumps contribute $\sim 0.5\% - 15\%$ of the
galaxy-integrated rest-frame $\approx 5000$\,\AA\ emission, with median
of $\approx 2\%$; the total contribution of clump light ranges from
$10\% - 25\%$.
The median intrinsic clump size and stellar mass are $\sim 1~{\rm kpc}$
and $\rm \sim 10^{9}~M_{\odot}$, in the ranges for clumps identified in
rest-UV or line emission in other studies.
The clump sizes and masses in the subset of disks are broadly consistent
with expectations for clump formation through gravitational instabilities
in gas-rich, turbulent disks given the host galaxies' global properties.
By combining the NIC2 data with ACS/F814W imaging available for one source,
and adaptive-optics assisted SINFONI H$\alpha$ data for another, we infer
modest color, $M/L$, and stellar age variations within each galaxy.
In these two objects, sets of clumps identified at different wavelengths do
not fully overlap; NIC2-identified clumps tend to be redder/older than ACS-
or H$\alpha$-identified clumps without rest-frame optical counterparts.
There is evidence for a systematic trend of older ages at smaller
galactocentric radii among the clumps, consistent with scenarios
where inward migration of clumps transports material towards the
central regions.  From constraints on a bulge-like component at radii
$\la 1 - 3~{\rm kpc}$, none of the five disks in our sample appears to
contain a compact massive stellar core, and we do not discern a trend
of bulge {\em stellar} mass fraction with stellar age of the galaxy.
Further observations are necessary to probe the build-up of stellar
bulges and the role of clumps in this process.

\end{abstract}


\section{INTRODUCTION}
         \label{Sect-intro}

Luminous, kiloparsec-sized ``clumps'' appear to be a ubiquitous feature
of high-redshift star-forming galaxies.
Such clumps have long been discussed in the context of high resolution
imaging of distant objects \citep*[e.g.,][]
{Cow95,vdB96,Col96,Gia96,Elm04a,Elm04b,Elm05a,Con04,Lot04,Pap05,Law07}.
They have been mostly identified in optical imaging from the
{\em Hubble Space Telescope\/} ({\em HST\/}), probing rest-frame UV
wavelengths at $z \ga 1$.  High resolution {\em HST\/} near-infrared
(near-IR) observations with the NICMOS camera, and most recently with the
WFC3 instrument, show that clumpy structures can also be present in the
rest-frame optical
\citep[e.g.,][]{Tof07, Das08, Kri09, Elm09a, Buss09, Swi10a, Ove10, Cam10}.
Similar clumps have been identified in rest-frame optical line emission
from near-IR integral field spectroscopy of several $z \sim 2$ disks
\citep{Gen06, Gen08, Gen11}.
Observations of strongly lensed $z \sim 1 - 3$ sources in broad-band
rest-UV/optical light, rest-optical line emission, and submillimeter
dust continuum and CO line emission with source-plane resolution down
to $\rm \sim 100~pc$ have confirmed characteristic sizes for stellar
clumps and star-forming complexes of $\rm \sim 300~pc - 1~kpc$
\citep[e.g.,][]{Sta08, Jon10, Swi10b}.

Various lines of evidence suggest that these clumps are not only
associated with distinct components in merging or interacting systems.
They are found in all types of galaxies at $z \ga 1$ including disks
identified as such on the basis of their structural properties
\citep{Elm05a, Elm05b, Elm07, Elm09a} as well as their kinematics
\citep{Gen06, Gen08, Gen11, Bour08, Jon10}.
High-redshift disks are found to have high gas to total baryonic mass fractions
of $\sim 20\% - 80\%$ and high local intrinsic gas velocity dispersions
of $\rm \sim 20 - 90~km\,s^{-1}$
\citep[e.g.,][]{Erb06,FS06,FS09,Gen08,Cre09,Epi09,Law09,Wri09,Dad10,Tac10}.
Numerical simulations of gas-rich turbulent disks indicate that massive
kpc-sized clumps can form in-situ through gravitational instabilities
\citep*[e.g.,][]
{Nog99,Imm04a,Imm04b,Bour07,Elm08,Dek09,Age09,Cev10,Aum10,Genel10}.
The Toomre length and mass derived from the global properties of clumpy
disks at high redshift correspond roughly to the clump sizes and masses
inferred from observations \citep*[e.g.,][]{Gen08,Gen11,Bour08,Elm09a,Jon10}.
In these numerical simulations, and according to theoretical arguments,
the clumps can migrate towards the gravitational center as a result of
their mutual interactions and of dynamical friction against the host disk,
and coalesce into a young bulge on timescales of $\rm \sim 1~Gyr$.  They
may also lose part of their mass, contributing to the redistribution of
mass and angular momentum and to the growth of the disk.

This ``clump-driven'' scenario has gained popularity and empirical support
in recent years, as an important mechanism for the early formation phases
of the spheroidal and disk components of present-day massive galaxies.
One of the major uncertainties in this scenario is the survival of clumps
against the disruptive effects of tidal torques and of stellar feedback,
both radiative and mechanical \citep{Mur10, Kru10, Genel10, Gen11}.
Characterizing in detail the properties of individual clumps in distant
galaxies, such as their masses, ages, star formation histories, and spatial
distribution across the host galaxy, is crucial to address this issue but
remains obviously very challenging because of the sensitivity and high
spatial resolution required.

In this paper, we study clumps identified in the {\em rest-frame optical\/}
emission in six massive star-forming $z = 2.1 - 2.5$ galaxies, for which we
obtained deep high-resolution {\em HST\/} imaging at 1.6\,\micron\ with the
NICMOS/NIC2 camera and F160W filter.  The targets are among the rest-UV
selected sources observed with SINFONI as part of the SINS survey
\citep{FS06, FS09}.
In a companion paper \citep[][hereafter Paper~I]{FS11}, we derived the
structural and morphological parameters of the targets from the NIC2
imaging, analyzed the results together with the H$\alpha$ line maps and
kinematics from SINFONI, and compared the morphological properties with
those of other samples of massive $z \sim 2$ galaxies imaged with NIC2/F160W
but selected very differently and probing a wide range in star formation
activity.  All but one of our SINS NIC2 targets exhibit compelling kinematic
signatures of disk rotation.  The global rest-frame optical surface
brightness distributions indicate disk-dominated morphologies as well,
with S\'ersic indices in the range $n \approx 0.15 - 2$.  In contrast,
the kinematics of the sixth source unambiguously reveal a major merger,
the main components of which are well resolved in the NIC2 imaging.
Here, we analyze the prominent kpc-scale structures seen in all our targets.
The combination of SINFONI and NIC2 data sets enables us to determine whether
observed clumps represent structures consistent with an in-situ formation
through disk instabilities or, instead, whether they are associated with
merging components.  We focus on the stellar properties of the clumps in
order to constrain their evolution and role in early bulge formation.
Two objects are studied in more detail, taking advantage of additional
high resolution {\em HST\/} optical imaging available for one, and
SINFONI observations together with adaptive optics (AO) for the other.

The paper is organized as follows.
We present the selection and global properties of the sample, and summarize
the NICMOS/NIC2 observations and data reduction in \S~\ref{Sect-data}.
We describe the identification procedure and derive the main properties
of clumps in the NIC2 images in \S~\ref{Sect-res}.
In \S~\ref{Sect-cases}, we focus on detailed case studies of the target
with existing high-resolution {\em HST\/} optical imaging (SSA22a$-$MD\,41)
and the one with AO-assisted H$\alpha$ SINFONI observations (Q2346$-$BX\,482),
and further constrain spatial variations in colors, stellar mass-to-light
ratios, and stellar ages.
The implications of our results for the formation and
evolution of clumps are discussed in \S~\ref{Sect-disc}.
The paper is summarized in \S~\ref{Sect-conclu}.
Throughout, we assume a $\Lambda$-dominated cosmology
with $H_{0} = 70\,h_{70}~{\rm km\,s^{-1}\,Mpc^{-1}}$,
$\Omega_{\rm m} = 0.3$, and $\Omega_{\Lambda} = 0.7$.
For this cosmology, 1\arcsec\ corresponds to $\rm \approx 8.2~kpc$
at $z = 2.2$.  Magnitudes are given in the AB photometric system
unless otherwise specified.

\section{SAMPLE, OBSERVATIONS, AND DATA REDUCTION}
         \label{Sect-data}

\subsection{NIC2 Target Selection and Ancillary Data}
            \label{Sub-sample}

Table~\ref{tab-targets} lists the six galaxies studied in this paper.
The targets were drawn from the initial sample of 17 rest-UV selected
objects at $z \sim 2$ from the SINS survey \citep{FS06, FS09} carried out
with the near-IR integral field spectrometer SINFONI \citep{Eis03, Bon04}
at the ESO VLT.  The galaxies were originally part of the large optical
spectroscopic survey of $z \sim 1.5 - 2.5$ candidates selected by their
$U_{n}G\mathcal{R}$ colors described by \citet{Ste04}.
Additional multi-wavelength data include ground-based near-IR $J$
and $K_{\rm s}$ band imaging, and space-based {\em Spitzer\/} mid-IR
photometry at $\rm 3 - 8~\mu m$ with IRAC and at 24\,\micron\ with MIPS
for the majority of our targets \citep{Erb06, Red10}.  For one galaxy,
MD\,41, high resolution optical imaging is available from observations
with the {\em HST\/} Advanced Camera for Surveys (ACS) through the F814W
filter (hereafter $i_{814}$ bandpass).

The choice of our NIC2 targets was primarily driven by their
kinematic nature along with their high signal-to-noise (S/N), high
quality SINFONI data mapping the spatial distribution and relative gas
motions from the H$\alpha$ line emission out to radii $\rm \ga 10~kpc$.
Five of the targets (BX\,663, MD\,41, BX\,389, BX\,610, and BX\,482)
were explicitly selected because of their disk-like kinematics, and the
sixth one (BX\,528) because it shows in contrast kinematic signatures
consistent with a counter-rotating binary merger.  The galaxies are
among the brighter and larger half of the full SINS sample in terms
of H$\alpha$ flux and size (see \citeauthor*{FS11}).  The quantitative
kinematic classification was performed through application of kinemetry
by \citet{Sha08}, and detailed dynamical modeling of the disks is presented
by \citet{Gen08} and \citet{Cre09}.

The SINFONI data sets of our NIC2
targets, among the deepest of the SINS survey, are fully described by
\citet{FS06, FS09}.  Briefly, all galaxies were observed in the $K$
band (targeting the H$\alpha$ and [\ion{N}{2}]\,$\lambda\lambda 6548,6584$
emission lines) using the largest pixel scale of $\rm 125\,mas\,pixel^{-1}$,
five of them in seeing-limited mode with FWHM spatial resolution of
$\approx 0\farcs 5$ (corresponding to $\rm \approx 4.1~kpc$) and one of
them, BX\,663, with the aid of AO resulting in a resolution of $0\farcs 39$
(or 3.2~kpc).  Seeing-limited $H$-band data at the $\rm 125\,mas\,pixel^{-1}$
scale were taken for all but BX\,528 to map the H$\beta$ and
[\ion{O}{3}]\,$\lambda\lambda 4959,5007$ emission lines, also under very
good near-IR seeing of $\approx 0\farcs 55$.  For BX\,482, AO-assisted
$K$ band observations were obtained using the intermediate pixel scale
of $\rm 50\,mas\,pixel^{-1}$, with total on-source integration time of
6.8\,hours and resolution of $0\farcs 17$ (1.4~kpc), very similar to
that of the NIC2 1.6\,\micron\ imaging.

\subsection{Global Properties of the NIC2 Targets}
            \label{Sub-sampleprop}

Our NIC2 sample is relatively bright in the near-IR.  The five objects
with existing $K$ band imaging have a median $K_{\rm s,AB} = 21.8~{\rm mag}$
and span a range of $\rm \approx 1.2~mag$ in observed $K$ band as well as in
rest-frame absolute $V$ band ($K_{\rm s,AB} = 21.1 - 22.3~{\rm mag}$, and
$M_{V,{\rm AB}}$ between $-23.1$ and $\rm -21.9~mag$; see \citeauthor*{FS11}).
{\em Spitzer\/}/MIPS 24\,\micron\ measurements that have been obtained for
four of the targets (BX\,389, BX\,610, BX\,528, and BX\,663) indicate that
they are among the 3\% most luminous at rest-frame 8\,\micron\ wavelengths
of UV-selected $z \sim 2$ galaxies \citep{Red10}.

In terms of global stellar properties, our NIC2 targets probe the
actively star-forming part of the massive $z \sim 2$ galaxy population
\citep{Erb06, FS09}.  Modeling of their optical to near-IR spectral
energy distributions (SEDs), including the new photometry obtained
with NIC2, is described in \citeauthor*{FS11}.  The results using the
\citet{BC03} evolutionary synthesis code, assuming constant star
formation, solar metallicity, and adopting the \citet{Chab03} IMF and
\citet{Cal00} reddening law, are given in Table~\ref{tab-targets}.
The stellar masses span over an order of magnitude, from
$7.7 \times 10^{9}$ to $\rm 1.0 \times 10^{11}~M_{\odot}$ with median
$\rm \approx 5 \times 10^{10}~M_{\odot}$.  The median absolute and specific
star formation rates (SFRs) are $\rm \approx 50~M_{\odot}\,yr^{-1}$ and
$\rm 0.7~Gyr^{-1}$ (with ranges of $\rm 25 - 185~M_{\odot}\,yr^{-1}$ and
$\rm 0.6 - 24~Gyr^{-1}$).  The objects are moderately obscured, with $A_{V}$
between 0.6 and 1.2~mag, or equivalently $E(B-V) = 0.15 - 0.30~{\rm mag}$.  
The stellar ages range from 50~Myr to 2.75~Gyr; BX\,528, BX\,663, BX\,389,
and BX\,610 have ages comparable to the age of the Universe at their
redshift, indicating that they host mature stellar populations.  
The best-fit models to the SEDs of these four objects imply rest-frame
$(U-B)_{\rm AB} \approx 0.75 - 0.9~{\rm mag}$ colors, close to or just
about at the separation adopted by, e.g., \citet{Kri09}, between ``red''
and ``blue'' objects.  In contrast, the two youngest objects, MD\,41
and BX\,482 are significantly bluer with rest-frame
$(U-B)_{\rm AB} \approx 0.5~{\rm mag}$.

Two of our targets (BX\,663 and BX\,610) are identified as candidate
AGN on the basis of their observed mid-IR SEDs from {\em Spitzer\/}
IRAC and MIPS photometry \citep{Red10}.  BX\,663 also exhibits spectral
signatures of Type 2 AGN in its integrated rest-frame UV and optical
spectra \citep{Sha04, Erb06}.  In our spatially-resolved SINFONI data,
the central compact emission peak of BX\,663 is characterized by a higher
[\ion{N}{2}]/H$\alpha$ and a broad H$\alpha$ velocity component associated
with the AGN, underneath the narrower component dominated by star formation.
In contrast, BX\,610 exhibits no sign of an AGN in the rest-frame UV and
optical, so that its putative AGN is likely to be very obscured and will
not affect any aspect of our analysis.  Alternatively, its mid-IR emission
properties could be due to enhanced PAH emission around rest-frame 8\,\micron\
with respect to its far-IR luminosity and in comparison to local star-forming
galaxies of the same total luminosity, as indicated by recent measurements of
the full mid- to far-IR SEDs of high-$z$ star-forming galaxies
\citep[e.g.][]{Nor10, Elb10, Muz10, Nor11, Wuy11}.
In the context of this paper, we will only consider BX\,663 explicitly
as an AGN source.

As described in \citeauthor*{FS11}, our NIC2 targets do not stand out
compared to the full SINS sample in terms of global stellar and dust
properties but they do more so in terms of their observed H$\alpha$
properties.
This is a consequence of our emphasis on selecting objects for the NIC2
follow-up with highest S/N and spatially best-resolved H$\alpha$ emission
from the initial seeing-limited SINFONI data sets from the SINS survey.
The median stellar mass and SFR of our NIC2 targets are comparable to those
of the SINS H$\alpha$ sample ($\approx 3 \times 10^{10}~{\rm M_{\odot}}$
and $\rm \approx 70~M_{\odot}\,yr^{-1}$).  However, they are among the
brighter half in terms of integrated H$\alpha$ fluxes (uncorrected for
extinction) with median 
$F({\rm H\alpha}) = 1.9 \times 10^{-16}~{\rm erg\,s^{-1}\,cm^{-2}}$
(and range from 1.1 to $\rm 3.1 \times 10^{-16}~erg\,s^{-1}\,cm^{-2}$),
compared to the SINS H$\alpha$ sample median of 
$F({\rm H\alpha}) = 1.1 \times 10^{-16}~{\rm erg\,s^{-1}\,cm^{-2}}$.
Our NIC2 targets also lie at the high end of the H$\alpha$ velocity-size
distribution \citep{Bou07, FS09}.  Their intrinsic half-light radii
$r_{1/2}{\rm (H\alpha)} \approx 4 - 5~{\rm kpc}$ are above the SINS
sample median of 3.1~kpc, and the circular velocities $v_{\rm d}$ for
the five disks are equal to or larger than the median for SINS galaxies
of $\rm \approx 180~km\,s^{-1}$, and up to $\rm \sim 300~km\,s^{-1}$
\footnote{
 The merger BX\,528 has a lower equivalent $v_{\rm d}$ of
 $\rm 145~km\,s^{-1}$; see \citet{FS09} for the derivation.}.
The ratio of circular velocity to intrinsic local velocity dispersion
of these disks are in the range $v_{\rm c} / \sigma_{0} \approx 2 - 6$
\citep{Gen08, Cre09, FS09}, which is significantly lower than for
local spiral galaxies
\citep*[with typical $v_{\rm c}/\sigma_{0} \sim 10-20$; e.g.,][]{Dib06,Epi10}.
These low ratios suggest comparatively larger gas turbulence and geometric
thickness, and appear to be a characteristic feature of early disk galaxies
at $z \sim 1 - 3$ \citep[e.g.][]{FS06,Wri07,Gen08,Sta08,Cre09,Epi09,Jon10}.

\subsection{NICMOS/NIC2 Observations and Data Reduction}
            \label{Sub-obs_red}

The spatial extent and brightness of our targets make them particularly
well suited for a detailed study of morphologies and a reliable assessment
of structural parameters from high resolution rest-frame optical continuum
imaging (\citeauthor*{FS11}).  Sensitivity and adequate sampling of the
instrumental point-spread function (PSF) are essential for these purposes,
as well as for identifying and characterizing small-scale structure, the
subject of this paper.  These goals dictated our observational strategy
and the procedures applied for data reduction, PSF characterization, and
analysis of the noise properties, which are detailed in \citeauthor*{FS11}
and summarized in this subsection.

The NICMOS observations were carried out between 2007 April and 2007
September with the NIC2 camera onboard the {\em HST\/} and using the F160W
filter with mean wavelength of 1.61\,\micron\ (hereafter $H_{160}$ bandpass).
NIC2 has a pixel scale of $0\farcs 075$, critically sampling the {\em HST\/}
PSF at 1.6\,\micron.
The $H_{160}$ bandpass probes rest-frame optical wavelengths around 5000\,\AA\
at the median $z = 2.23$ of our targets, and corresponds closely to the SDSS $g$
band (which is about 300\,\AA\ bluer in the rest-frame).  At these wavelengths,
redward of the 4000\,\AA/Balmer break, the galaxies' light traces the bulk of
stellar mass more reliably than the rest-frame UV, which is significantly
more affected by recent star formation and by dust extinction.  Each target
was observed for four orbits, amounting to a total on-source integration time
of 10240\,s with good sampling of the PSF.

The data were reduced using standard routines of the NICMOS pipeline
within the IRAF environment, complemented with custom procedures to optimize
the removal of bad pixels and cosmic ray hits, and of large scale residuals
across the images.  The images were drizzled onto a final pixel scale of
$\rm 0\farcs 05~pixel^{-1}$.  The PSF was determined empirically from the
two-dimensional profiles of the four brightest, unsaturated, and isolated
stars present in our NIC2 pointings, and has a FWHM of $0\farcs 145$ based
on a Gaussian profile fit.  The effective noise properties were characterized
from the final reduced images, taking into account deviations from pure
Gaussian noise due to instrumental features as well as correlated noise
and systematics resulting from the data reduction procedure.
The $3\,\sigma$ limiting surface brightness of the final
images is $\mu(H_{\rm 160,AB}) = 23.4~{\rm mag~arcsec^{-2}}$, and the
$3\,\sigma$ limiting magnitude in a ``point-source aperture'' with
diameter $d = 1.5 \times {\rm PSF~FWHM} = 0\farcs 22$ (maximizing the S/N
of photometric measurements in unweighted circular apertures of point-like
sources) is $\rm \approx 28.1~mag$.  The total $H_{160}$ band magnitudes
measured in a circular aperture with diameter of 3\arcsec\ centered on
each target are reported in Table~\ref{tab-targets}.

Figure~\ref{fig-clumpsid} presents the final reduced images of the
galaxies with sampling of $\rm 0\farcs 05~pixel^{-1}$.  The angular
resolution of $\rm FWHM = 0\farcs 145$ corresponds to a spatial
resolution of $\rm \approx 1.2~kpc$ at the redshifts of our targets.
The geometric center calculated as the unweighted mean of the $x$ and $y$
coordinates of pixels with $\rm S/N \geq 3$ is indicated by a cross in
Figure~\ref{fig-clumpsid}.  For the five disks, it corresponds closely
to the dynamical center based on modeling of the H$\alpha$ kinematics
\citep{Gen08, Cre09}.
In $H_{160}$ band (rest-frame $\approx 5000$\,\AA) emission, the galaxies
exhibit a diffuse and lower-surface-brightness component extending out to
radii of $0\farcs 8 - 1\farcs 2$, or $\rm 6.5 - 10~kpc$.  Rich substructure,
including several bright compact emission regions, is seen in all sources
and contributes to their irregular appearance.
For BX\,528, two major emission sources to the southeast and northwest
are well resolved in the NIC2 images (denoted BX\,528$-$SE and NW, with
projected separation of 8~kpc), with relative $H_{160}$ band fluxes that
suggest a nearly equal-mass major merger (mass ratio of $\sim 1.5 : 1$),
in agreement with the kinemetry analysis of the H$\alpha$ kinematics
\citep{Sha08}.  For the large disk, BX\,389, the small and faint companion
$\rm 5~kpc$ in projection to the south of, and at nearly the same redshift
as the main part of the galaxy signals a minor merger (mass ratio of
$\sim 10 : 1$), consistent with the lack of substantial disturbances
in the H$\alpha$ kinematics.

\section{PROPERTIES OF CLUMPS IDENTIFIED IN THE REST-FRAME OPTICAL}
         \label{Sect-res}

In \citeauthor*{FS11}, we quantified the overall structural properties 
of our NIC2 targets using both parametric and non-parametric methods.
We compared the rest-frame optical and H$\alpha$ morphologies (as well
as the rest-frame UV morphology for MD\,41), and also made a differential
comparison with other samples of massive galaxies at the same redshift
but selected using different photometric criteria.
The effective radius $R_{\rm e}$, S\'ersic index $n$, and ratio of minor
to major axis $b/a$ derived from single-component S\'ersic model fits to
the $H_{160}$ band maps of our targets are given in Table~\ref{tab-targets}.
Now, we turn to the prominent and widely distributed clumpy features that
constitute one of the most striking aspects of the rest-frame optical
morphologies of our SINS NIC2 galaxies.  In what follows, we use the
term ``clump'' to refer generically to any small-scale feature identified
with the method described below.  This method also recovers the merger
components of BX\,528 as clumps, as well as the southern companion of
BX\,389 and the central peak of the AGN-hosting BX\,663.

\subsection{Clump Identification and Measurements}
            \label{Sub-clumps_meth}

To identify clumps as objectively as possible, and in the same way for all our
galaxies, we used the IRAF task {\em daofind\/} \citep{Ste87}.  This task is
optimized for identification of compact or point-like sources in crowded
fields with variable underlying background emission, and is thus appropriate
for our purposes.  This task detects objects by searching for local density
maxima with specified FWHM and peak amplitudes above the local background.
We used a detection threshold of $\geq 3\,\sigma$ (with $\sigma$ being
the pixel-to-pixel rms noise) and, given the observed extent of the bright
clumps, we set the characteristic scales to $\rm 1 - 1.5~\times$ the PSF
FWHM.  Because the clumps are not strictly point sources and have a range
of sizes, brightnesses, and brightness contrasts against the host galaxies'
``background,'' the number of clumps depends slightly on the exact values
for the threshold and size scale.  Based on experimentation, we found that
the choices above were the most satisfactory according to a visual
assessment, and we used the same values for all galaxies.
A total of 28 clumps are identified, with between 2 and 7 per source,
and are indicated on the NIC2 maps in Figure~\ref{fig-clumpsid}.
They are listed in Table~\ref{tab-clumpsprop} along with their projected
distance $d_{\rm proj}$ from the geometric center of the galaxies
(coinciding with the dynamical center) and other properties derived below.
We also calculated the deprojected
galactocentric distance $d$ assuming the clumps are confined to a disk
of inclination with respect to the sky plane given by the best-fit
morphological axis ratio of each galaxy (Table~\ref{tab-targets}).
As seen in Figure~\ref{fig-clumpsid}, the two main components of the
interacting system BX\,528 correspond to the neighbouring clumps \#1
and \#2 (BX\,528$-$SE) and to clump \#4 (BX\,528$-$NW).  The central
peak of BX\,663 is identified as clump \#2, and the southern companion
of BX\,389 as clump \#4.

Measurements of the clump sizes and brightnesses in our NIC2 images are
complicated by the bright and variable background from the host galaxy,
and by the close proximity of neighboring clumps often with relatively
few resolution elements sampling both clumps and interclump regions.
Moreover, large-scale asymmetries in the global light distribution of
some of the galaxies affect the accuracy of the clump properties if using
the residual maps from the S\'ersic model fits (see \citeauthor*{FS11}).
The light profiles extracted in circular apertures of increasing radius
centered on the clump positions generally show an upturn at approximately
mid-distance between bright adjacent clumps.  For more isolated clumps or
for those next to other comparatively fainter clumps, the profiles decrease
monotonically but show a break where they become dominated by the underlying
galaxy light or where the fainter neighboring clumps start to contribute
more significantly.  In all those cases, we took the average brightness
at the upturn or break of the profiles in annuli of width
$dr = 1 - 1.5~{\rm pixels}$ to represent the local background, and computed
the background-subtracted clump brightnesses within the apertures of radius
$r_{\rm phot}$ just inside these annuli.
For the other few clumps with no upturn or clear break in their profile,
we adjusted $r_{\rm phot}$ so as to maximize the aperture size without
overlapping with those of neighboring clumps, with the background also
measured in narrow annuli just outside of the clump photometric apertures.

The photometric apertures have a diameter $d_{\rm phot}$ in
the range $\approx 1.4 - 3.8 \times$ the PSF FWHM, on average $2 \times$,
and would typically miss a significant fraction of the total flux even for
compact or point-like sources.  Since the observed size of the clumps is 
comparable to the spatial resolution and their intrinsic shape is unknown,
we applied a correction factor equal to the ratio of total flux (enclosed
within a radius of 1\arcsec) and the flux within $r_{\rm phot}$ based on
the PSF curve-of-growth.  The average correction is a factor of 1.7, with
a range from 1.14 to 2.09.
The background measured over $r_{\rm phot} < r < r_{\rm phot} + dr$ also
includes a contribution from the clump itself, which is, for the narrow
annuli used, $10\%$ on average (and in the range $4\% - 14\%$) of the
total flux for a point source.  This fraction is small compared to both
the aperture correction and the background estimate, so we neglected this
additional correction.  We determined clump sizes by taking the direct
FWHM of the background-subtracted light profiles (i.e., twice the radius
at which the profile is half its maximum value).  Nine of the 28 clumps
are unresolved, with direct FWHM smaller than the PSF FWHM; we adopted the
PSF FWHM as an upper limit on their size.  For the others, the intrinsic
FWHM size is simply calculated by subtracting in quadrature the observed
clump and PSF FWHMs.  The FWHM sizes, $H_{160}$ band magnitudes, and
fractions of the total galaxy light contributed by the clumps (denoted
$\rm FWHM^{cl}$, $H_{160}^{\rm cl}$, and $\mathcal{F}^{\rm cl}$) as
well as the photometric aperture radii and corrections are given in
Table~\ref{tab-clumpsprop}.  

To gauge possible uncertainties on the clump light contributions, we
considered two other measurement methods.  In the first, we used the peak
flux above the local background as returned by {\em daofind} when identifying
the clumps and computed the total flux assuming Gaussian unresolved sources.
Since the clumps are mostly ($\sim 2/3$) consistent with being extended ---
albeit compact --- sources, this method is expected to provide lower limits.
The inferred fluxes are on average (and median) a factor of $\approx 3$
lower than the adopted estimates described just above, and with smaller
differences typically found for the unresolved clumps (with $\rm FWHM^{cl}$
$<$ PSF FWHM).  Alternatively, we estimated the integrated light within
apertures of radius $r_{\rm phot}$ but without background subtraction and
aperture correction, in a manner analogous to that used by \citet{Elm09a}.
In view of the significant surface brightness of the
underlying light in our galaxies, these estimates represent upper limits
and are typically a factor $\approx 2.5$ higher than the adopted values.
The corresponding fractional light contributions obtained with these
two approaches ($\mathcal{F}^{\rm cl}_{\rm PS}$ and
$\mathcal{F}^{\rm cl}_{\rm raw}$, respectively)
are also given in Table~\ref{tab-clumpsprop} for comparison with the
adopted measurements.  We thus infer that the adopted clump light
contributions are uncertain to a factor of $\approx 3$.

We emphasize that the contrast between the clumps and the surrounding,
or interclump regions in the $H_{160}$ band images of our NIC2 targets
is relatively small.  In the photometric apertures employed, the total
contribution from the local background light is typically $\sim 3 - 4$
times higher than the background-subtracted clump fluxes, making the
measurements of the clump brightnesses rather sensitive to the background
estimates as seen above.
This is very different from the ``clump-cluster,'' ``chain,'' and spiral
galaxies from the Hubble Ultra Deep Field (H-UDF) studied in the optical
by \citet{Elm05a} and \citet{Elm09a}, in which the clumps identified in
ACS F775W imaging ($i_{775}$ band) are typically brighter by factors of
$\sim 2 - 4$ than the surrounding emission, with the contrast increasing
further at shorter wavelengths.
While \citet{Elm05a} accounted for the galaxy's background in measuring
clump magnitudes, \citet{Elm09a} neglected this contribution since it is
generally small in the optical for their sample of H-UDF galaxies.

We computed the approximate observed (i.e., uncorrected for dust
extinction) rest-frame SDSS $g$ band luminosities of the clumps
from the background-subtracted and aperture-corrected $H_{160}$ band
magnitudes, adopting, to express the results in solar units, an absolute
$M_{g,\odot} = 5.07~{\rm mag}$ \citep{Bla01}\,\footnote{
 The NIC2 $H_{160}$ bandpass probes a wavelength range between the
 rest-frame $g$ and $V$ bands for our targets, with closer effective
 wavelength and larger overlap with the $g$ bandpass.  For the best-fit
 SED parameters of the sources, the color term between observed $g$ and
 $V$ band magnitudes, or luminosities, varies from $3\%$ to $20\%$ depending
 on the object, with median/mean of $\approx 10\%$; the difference between
 the actual rest-frame luminosity probed through the $H_{160}$ bandpass and
 the $g$ band would be smaller.}.
We further estimated the clump stellar masses by multiplying the $M_{\star}$
derived from the SED modeling of each galaxy with $\mathcal{F}^{\rm cl}$.
This scaling corresponds to assuming, crudely, a constant ratio across
each source between stellar mass and observed $H_{160}$ band light or,
equivalently, rest-frame $g$ band light.
For MD\,41, the $M_{\star}/L_{g}^{\rm rest}$ ratio map obtained from the
$i_{814} - H_{160}$ color map constructed from its NIC2 and ACS images
indicates a fairly small scatter of $\rm \approx 0.2~dex$ in observed
$\log(M_{\star}/L_{g}^{\rm rest})$ across the galaxy (see \citeauthor*{FS11}).
For the clumps themselves, the more detailed analysis presented in
\S~\ref{Sect-cases} for MD\,41 and BX\,482 suggests that the assumption of
constant $M/L$ ratio is on average valid to within a factor of $\sim 2 - 3$.
The clump luminosities and stellar masses, $L_{g}^{\rm rest,cl}$ and
$M_{\star}^{\rm cl}$, are listed in Table~\ref{tab-clumpsprop}.

\subsection{Clump Properties for the NIC2 Targets}
            \label{Sub-clumps_res}

Figure~\ref{fig-clumpsprop} plots the $H_{160}$ band magnitude, fractional
light contribution, and inferred stellar mass of the clumps as a function
of observed size.  The distribution of intrinsic sizes and stellar masses
of the resolved clumps is shown in Figure~\ref{fig-clumpsprop2}.
The observed FWHM sizes range from unresolved up to $\rm \approx 2~kpc$.
Excluding the nine unresolved clumps, the intrinsic $\rm FWHM^{0}$ sizes
span $\rm \approx 0.1 - 1.5~kpc$, with mean and median of
$\rm \approx 0.9~kpc$, or $\approx 0\farcs 11$.
The clump magnitudes vary between $H_{\rm 160,AB}^{\rm cl} = 24.3$ and
27.9~mag, with mean and median of $\rm \approx 26.5~mag$.  On average,
the clumps have $\mathcal{F}^{\rm cl} \approx 4\%$ (median $\approx 2\%$),
ranging from approximately $0.5\%$ to $16\%$.
With the constant $M/L$ ratio assumption for each source,
these fractions imply $M_{\star}^{\rm cl} \sim 9 \times 10^{7}$ to
$\rm 9 \times 10^{9}~M_{\odot}$, and typically $\rm \sim 10^{9}~M_{\odot}$.
Considering only the resolved clumps, the mean and median
$M_{\star}^{\rm cl} \sim 2 \times 10^{9}~{\rm M_{\odot}}$.
Among the resolved clumps, we find no significant correlation
between clump magnitude or fractional light contribution and size,
and the scatter of the data is fairly large ($\rm \approx 0.9~mag$
in $H_{160}$, and about equal to the mean in $\mathcal{F}^{\rm cl}$).
There is also no obvious trend of clump size, magnitude, fractional light
contribution, or stellar mass as a function of galactocentric distance
($d_{\rm proj}$ or $d$) nor of distance normalized to the $R_{\rm eff}$
of the galaxies.
In Figure~\ref{fig-clumpsprop}, panel {\em a\/}, dotted curves indicate
constant apparent central surface brightnesses, i.e., within a circular
aperture of diameter equal to the observed $\rm FWHM^{cl}$.
The clumps cover about 2 magnitudes in central surface brightness, with 
a standard deviation of $\rm \approx 0.7~mag~arcsec^{-2}$ around the
median (and mean) value of $\rm \mu_{160,AB} \approx 23.7~mag~arcsec^{-2}$.
This value corresponds roughly to the $3\,\sigma$ surface
brightness limit of our NIC2 images.

The ``clumps'' associated with the merger components of BX\,528, the southern
companion of BX\,389, and the central core of BX\,663 (plotted with different 
symbols in Figures~\ref{fig-clumpsprop} and \ref{fig-clumpsprop2}) do not
differentiate in terms of size but are part of the brighter half and have
$\mathcal{F}^{\rm cl}$ above the median value.  Excluding them from our
clump sample does not, however, significantly affect the ranges and the
mean or median properties inferred above nor the absence of correlation
between clump properties.

Despite being a striking feature in the appearance of all our targets,
our results imply that the individual clumps contribute typically only a
few percent of the total light.  Even with the substantial uncertainties
of a factor of $\approx 3$, each clump still represents a fairly small
contribution.  Summed up over all clumps identified in a given galaxy,
the total contributions are typically $\sim 15\%$ per object, ranging
from roughly $10\%$ to $25\%$ (at least $\sim 5\%$ from the lower limits
limits obtained in the  alternative ``point-source'' estimates made above,
and at most $\sim 50\%$ from the upper limits derived without subtracting
the host background in the photometric apertures).  Thus, for our targets,
the bulk of the rest-frame $\rm \approx 5000$\,\AA\ light appears to be
dominated by an underlying component from a more smoothly distributed
stellar population.  Alternatively, fainter, unresolved clumps and
clusters may also contribute to this extended emission component.

We searched for correlations between clump properties and the global
properties derived from the SED modeling of the host source.  There is no
significant correlation between clump size, magnitude, or fractional light
contribution with galaxy stellar mass, age, absolute and specific SFR,
nor between clump size and fractional light contribution and extinction.
The only possible trend is of fainter clump $H_{160}$-band magnitude with
higher $A_{V}$, with Spearman's rank correlation coefficient $\rho = 0.53$
and $2.8\,\sigma$ significance level
\footnote{
 The significance quoted corresponds to the number of standard
 deviations by which the sum-squared difference of ranks deviates
 from the expected value under the null hypothesis of statistical
 independence ($\rho = 0$).}.
The trend weakens when excluding
the clumps associated with the merging components of BX\,528, the southern
companion of BX\,389, and the central peak of the AGN source BX\,663
($\rho = 0.37$, $1.7\,\sigma$).  There is no apparent trend with
morphological parameters although it is noteworthy that BX\,610 and
BX\,482, the two disks with qualitatively most prominent clumps, have
higher Gini coefficients than MD\,41 and BX\,389, the two disks where
the clumps appear visually less regular in shape.  MD\,41 and BX\,389
are the most edge-on disks based on their axis ratio and have the
largest inferred $A_{V}$, so that higher and possibly patchy extinction
may affect the appearance of these galaxies more than for the others.
Our sample is small, however, and covers a limited range in host
galaxy parameters (\S~\ref{Sub-sampleprop}), so that it would be
difficult to discern all but the strongest possible trends.

\subsection{Constraints on Emission Line Contributions}
           \label{Sub-linecontrib}

At the redshift of our targets, the NIC2 F160W bandpass includes
the [\ion{O}{3}]\,$\lambda\lambda\,4959,5007$ doublet and H$\beta$.
It is thus important to assess whether the observed $H_{160}$ band
morphologies are biased by nebular line emission.
For all sources except BX\,528, SINFONI seeing-limited observations
of these lines with the $H$ band grating were taken as part of the SINS
survey \citep{FS09}, and are analyzed in detail by \citet{Bus11}.  The
[OIII] lines are detected in all galaxies except BX\,663, and H$\beta$ is
additionally detected in MD\,41 and BX\,610.  Some of the non-detections
can be attributed to the galaxies' lines being redshifted to the wavelength
of bright telluric lines, in particular H$\beta$ for BX\,663 and BX\,482.
Based on the integrated line fluxes (or the $3\,\sigma$ upper limits thereof),
we concluded in \citeauthor*{FS11} that the broad-band $H_{160}$ emission of
our NIC2 targets is dominated by the continuum, with only a modest to very
small fraction from the [\ion{O}{3}] and H$\beta$ lines.  The largest line
contribution is inferred for BX\,389, where the [\ion{O}{3}] lines make up
$24\%$ of the integrated $H_{160}$ band flux and, with the $3\,\sigma$ upper
limit on H$\beta$, the total line contribution is estimated at $< 29\%$.
BX\,610 has the smallest contributions, amounting to $\approx 6\%$ for all
three lines combined.  Our examination of the sources on a pixel-to-pixel
basis further indicates no important spatial variations in the emission
line contributions.

In the context of this paper, it is worth considering more specifically
whether the clumps identified in \S~\ref{Sub-clumps_meth} could be due
to nebular line emission rather than stellar continuum emission, since
their contribution also represents a modest fraction of the total $H_{160}$
band light.  A detailed assessment would require AO-assisted SINFONI $H$
band observations, but the existing seeing-limited data provide useful
indications.  For BX\,610 and BX\,482, the total fraction of light from
the identified clumps ($16\%$ and $25\%$, respectively;
Table~\ref{tab-clumpsprop}) is at least twice as high as the total emission
line contribution ($6\%$ and $< 12\%$; \citeauthor*{FS11}), implying a
major contribution from the continuum in these two sources.  For BX\,663,
MD\,41, and BX\,389, the line contributions (or $3\,\sigma$ upper limits
thereof) are comparable to or higher than the sum of the clump contributions
so that it is not possible to set the same constraints with the current data.
For BX\,482, the spatial distributions in H$\alpha$ and $H_{160}$ band follow
each other fairly well, albeit with some small differences discussed in the
next section.  Our analysis in \S~\ref{Sect-cases} implies that the H$\alpha$
to $H_{160}$ band flux ratio at the location of the clumps identified in the
NIC2 image is a factor of $\la 2$ the source-integrated ratio.  If we assume
approximately constant observed emission line ratios across the galaxy,
this would imply [\ion{O}{3}] and H$\beta$ contributions to the $H_{160}$
band emission of at most twice that of the global contribution, i.e.,
$\la 25\%$ (\citeauthor*{FS11}).  Again, this means the clumps need to
have a major component of continuum light in $H_{160}$ band.  From the
estimates above, and at least for BX\,610 and BX\,482, the clumps in the
NIC2 images do not appear predominantly caused by line emission.  More
generally, clumps identified in high resolution optical and near-IR
broad-band imaging of other samples \citep[e.g.,][]{Elm05a, Elm09a}
are ubiquitous in $z \ga 1$ galaxies, over a range of redshift that
precludes an origin entirely from emission lines in all cases.

\section{DETAILED CASE STUDIES: MD\,41 AND BX\,482}
         \label{Sect-cases}

The sizes, luminosities, and stellar masses derived in the previous section
reflect the current state of the clumps but provide little information on
their evolution.  In addition, the stellar masses were based on the crude
assumption of constant $M/L$ ratio across the galaxies.  In the following,
we examine two of our targets with existing high spatial resolution ancillary
data to assess possible variations in $M/L$ ratios and search for differences
in age-sensitive properties among clumps, and between clump and interclump
regions.  We combine our NIC2 maps with the available ACS imaging for MD\,41
to look at the $i_{814} - H_{160}$ colors, and with the AO-assisted SINFONI
H$\alpha$ map of BX\,482 to determine the H$\alpha$ equivalent widths.
As we will see, the $i_{814} - H_{160}$ colors provide notably tighter
constraints on the $M/L$ ratios, irrespective of dust extinction, stellar
age, and star formation history.  On the other hand, the H$\alpha$ equivalent
widths are insensitive to extinction if the line and continuum emission
originates from the same regions, and are thus most useful to constrain
variations in stellar ages.

\subsection{MD\,41: Constraints from NIC2 and ACS Observations}
            \label{Sub-case_md41}

\subsubsection{Rest-frame Optical- and UV-Identified Clumps}
               \label{Sub-case_md41_clumpsid}

The ACS $i_{814}$ band map of MD\,41 is shown in Figure~\ref{fig-md41_panel}
alongside the NIC2 $H_{160}$ band map and a color-composite image.  For the
analysis, the ACS map has been convolved to the spatial resolution of the NIC2
data and registered onto the same astrometric frame (see \citeauthor*{FS11}
for details).  The appearance of MD\,41 is broadly similar in $i_{814}$ and
$H_{160}$ band.  The measurements in \citeauthor*{FS11} of the structural
and morphological parameters yield comparable effective radius, S\'ersic $n$
index, axis ratio, as well as Gini $G$, Multiplicity $\Psi$, and $M_{20}$
coefficients for both images.  The $i_{814} - H_{160}$ colors across MD\,41
are also fairly uniform, with median of 0.95~mag and standard deviation of
0.37~mag over the regions where pixels have $\rm S/N > 3$ in both bands.
These results indicate that there is no strong $k$-correction between
the rest-frame UV and optical morphologies for MD\,41, consistent with the
general findings for rest-UV selected samples at $z \sim 2$, and in contrast
to those for samples selected at longer wavelengths with typically redder
rest-frame optical colors \citep{Pet07, Law07, Tof07, Cam10}.  

However, it is also evident from Figure~\ref{fig-md41_panel} that the
brightest peaks do not all coincide spatially between the two maps.
Therefore, we applied the procedure described in \S~\ref{Sub-clumps_meth}
to detect clumps in the ACS image matched to the PSF of the NIC2 data,
using the same detection parameters in the IRAF task {\em daophot}.
Four clumps are thus identified in $i_{814}$ band emission, one of which
corresponds to NIC2 clump \#4 whereas the three others have no counterpart
among the NIC2 clumps (hereafter referred to as ``ACS-\#1,'' ``ACS-\#2,''
and ``ACS-\#3'').  All of the clumps are indicated on the maps of
Figure~\ref{fig-md41_panel}.  We computed the clump photometry with
background subtraction and aperture correction, as well as without for
comparison purposes (hereafter simply ``background-subtracted'' and ``raw''
photometry).  We measured the $i_{814}$ band brightness of the NIC2 clumps
in the same apertures used to derive their $H_{160}$ band brightnesses.
For the three additional ACS clumps, we defined appropriate apertures
as explained in \S~\ref{Sub-clumps_meth} but based on the $i_{814}$ band
light profiles.  The ACS-only clumps have no local excess emission above
our detection threshold in the NIC2 map, so we adopted $3\,\sigma$ upper
limits on their background-subtracted $H_{160}$ band magnitudes based on
the noise properties of the image.
For the NIC2-identified clumps, the background-subtracted $i_{814}$ band
magnitudes are formally above the $3\,\sigma$ noise level in the photometric
apertures except for clump \#1, for which we used a $3\,\sigma$ upper limit.

\subsubsection{Variations in Colors of the Clumps}
               \label{Sub-case_md41_colvar}

The photometric measurements are reported in
Table~\ref{tab-clumpsprop_md41} and plotted in Figure~\ref{fig-md41_pixplots}.
Panel {\em (a)\/} shows the $H_{160}$ versus $i_{814}$ band magnitudes for
individual pixels with $\rm S/N > 3$ in each band.
Panel {\em (b)\/} shows the magnitudes estimated for the clumps,
with and without subtraction of the galaxy's background.  For reference, the
total magnitudes of MD\,41 (in a circular aperture of 3\arcsec\ diameter) and
those corresponding to the sum of the fluxes of all pixels with $\rm S/N > 3$
in both bands are shown as well\,\footnote{
  The small difference between these two
 ``integrated'' measurements is due to the different areas sampled, with the
 total aperture including the two sources $\sim 1^{\prime\prime}$ southeast
 and northwest of MD\,41 seen in the ACS image but undetected in the NIC2
 image.}.
In both panels of Figure~\ref{fig-md41_pixplots}, the diagonal lines
correspond to constant $i_{814} - H_{160}$ colors.  A vector shows the
significant impact of extinction in the rest-frame of the galaxy on the
observed colors.  

All the NIC2 and ACS
clumps in MD\,41 contribute small fractions of $\sim 1\% - 5\%$ to the
galaxy-integrated fluxes in both $H_{160}$ and $i_{814}$ band emission
(rest-frame $\approx 5000$\,\AA\ and $\approx 2600$\,\AA), respectively.
Overall, the clump and interclump regions in MD\,41 span a comparable range
in their colors, reflecting the uniformity across the galaxy noted above.
The median color of the interclump regions is 1.0~mag, nearly equal to the
median over all $\rm S/N > 3$ pixels (0.95~mag).  The colors of the clumps
are mostly within $\rm \approx 0.5~mag$ of these values, for both the raw
and background-subtracted estimates (or limits).  Interestingly, there
appears to be a systematic difference in colors between the clumps identified
in the different bands: the NIC2 clumps have a median color roughly 0.6~mag
redder than the ACS clumps.  The two most distinct clumps are NIC2-\#1 and
ACS-\#3, which have colors up to 2~mag redder and bluer, respectively, than
the ensemble of the other clumps and the interclump regions.

\subsubsection{Implications for the Stellar Properties of the Clumps}
               \label{Sub-case_md41_impl}

With the $i_{814} - H_{160}$ colors of the clumps in MD\,41, we can
assess the reliability of the assumption made in \S~\ref{Sub-clumps_meth}
on their $M/L$ ratio and the consequences on the clump stellar masses.
To interpret the colors, we used the mean $i_{814} - H_{160}$ versus
$M_{\star}/L_{g}^{\rm rest}$ relationship derived in \citeauthor*{FS11}
from the evolutionary synthesis models of \citet{BC03} for a wide range
of star formation histories (SFHs), stellar ages, and dust extinction.
The observed $i_{814} - H_{160}$ colors are strongly degenerate in age,
extinction, and SFH.  On the other hand, the various model curves
occupy a fairly well defined locus in the $i_{814} - H_{160}$ versus
$M_{\star}/L_{g}^{\rm rest}$ parameter space, and the standard deviation
of the models suggest the relation is accurate to 0.15~dex in
$\log(M_{\star}/L_{g}^{\rm rest})$ over the range of colors observed across
MD\,41, with little sensitivity to metallicity between solar and $1/5$ solar
\citep[appropriate for our NIC2 targets including MD\,41;][]{Bus11}.
Stellar masses were then estimated by multiplying the $M/L$ ratios with
the observed luminosities from the $H_{160}$ band data.  The color-based
$M_{\star}/L_{g}^{\rm rest}$ and $M_{\star}$ estimates for all clumps
identified in MD\,41 are given in Table~\ref{tab-clumpsprop_md41}.

The color distribution of individual pixels across MD\,41 implies a median 
$\log(M_{\star}/L_{g}^{\rm rest}\,[{\rm M_{\odot}\,L_{\odot,g}^{-1}}])
 \approx -0.8$ over the interclump regions ($-0.85$ over all $\rm S/N > 3$
pixels).  For most clumps, the ratios (or limits thereof) derived from
the colors agree within a factor of $\sim 2 - 3$ with the median over
the interclump regions.  The most notable differences are for the reddest
and bluest clumps, NIC2-\#1 and ACS-\#3.  The ratio between the 
$M_{\star}/L_{g}^{\rm rest}$ values (or limits thereof) for these two
clumps ranges from $\approx 8 - 65$, depending on the treatment of
the galaxy's background.
The systematic offset in colors between the NIC2- and ACS-identified
clumps implies overall higher $M_{\star}/L_{g}^{\rm rest}$ ratios and
larger $M_{\star}$ for the former set of clumps by factors of $\sim 2$
and $\sim 4 - 6$, respectively (or more considering the limits).
For the NIC2-identified clumps \#2, \#3, and \#4, the background-subtracted
color-based masses inferred here are $\approx 1.5 - 2.5$ times larger than
those calculated in \S~\ref{Sub-clumps_meth} with the assumption that clumps
have the same $M/L$ ratio as the entire galaxy; for clump \#1, it is almost
a factor of eight higher.
The comparison of the trends based on the raw versus background-subtracted
photometry in MD\,41 indicates that, while still fairly modest between most
clumps, the variations in colors, hence $M/L$ ratios tend to be accentuated
(along with a general shift towards redder colors and higher $M/L$ ratios)
when accounting for the galaxy's background light contribution.

The observed distributions of colors in MD\,41 could be interpreted as
indicative of overall modest variations in extinction and ages between
clumps and across the galaxy, with the apparent exceptions of NIC2-\#1
and ACS-\#3.  The difference in median $i_{814} - H_{160}$ color between
the NIC2- and ACS-identified clumps by 0.6~mag could be accounted for
by a difference in $A_{V}$ of about 0.7~mag.  Relative to the interclump
regions, the redder color of NIC2-\#1 and bluer color of ACS-\#3 by up to
$\rm \approx 2~mag$ would require differences in $A_{V}$ of up to about
2.5~mag; given the best-fit $A_{V} = 1.2~{\rm mag}$ for the integrated
SED of MD\,41 (Table~\ref{tab-targets}), it appears plausible that not
only extinction but also age variations contribute to the observed clump
colors, at least for these two clumps.  If interpreted solely in terms
of stellar age, assuming constant star formation (CSF) and a uniform
$A_{V} = 1.2~{\rm mag}$, the NIC2-only clumps would be on average
$\sim 5 - 10$ times older than the ACS-only clumps.
Shorter timescales for the star formation activity would tend to reduce
(but not eliminate) the overall relative age difference between NIC2-
and ACS-only clumps, if extinction variations are neglected.

The results for MD\,41 suggest that, with the possible exception of the reddest
and bluest clumps, the $M/L$ and stellar masses are reliable to a factor of
a few.  The NIC2-identified clumps tend to be redder and more massive than
the ACS-identified clumps.  We find indications that not only extinction
effects but also variations in stellar age are likely responsible for the
full range in colors of the clumps.
The main uncertainties affecting our analysis are associated with
the limited photometry available for MD\,41 and with the difficulties
in evaluating the clump light in the presence of a bright background from
the host galaxy.

\subsection{BX\,482: Constraints from NIC2 and SINFONI Observations}
            \label{Sub-case_bx482}

\subsubsection{Rest-frame Optical- and H$\alpha$-Identified Clumps}
               \label{Sub-case_bx482_clumpsid}

For BX\,482, we can set more robust constraints on relative stellar ages
between the clumps and the interclump regions with the help of the SINFONI
AO-assisted H$\alpha$ map available for this galaxy.  Ideally, the H$\alpha$
line flux combined with the continuum emission measured in the near vicinity
of H$\alpha$ in wavelength would provide the H$\alpha$ equivalent width,
which varies strongly with the age of a stellar population but is insensitive 
to dust extinction\,\footnote{
 This statement is true unless different amounts of dust obscure the
 \ion{H}{2} regions compared to the bulk of the stellar population,
 or a significant fraction of Lyman continuum photons is absorbed by
 dust within the \ion{H}{2} regions or escapes from the \ion{H}{2}
 regions.  In the analysis presented here, we neglected these effects.
 There is evidence for such ``differential attenuation'' on global
 galactic scales locally \citep[e.g.,][]{Cal00} and at high $z$
 \citep[e.g.,][]{FS09, Man11, Yos11} but if it is similar at different
 locations across galaxies, the {\em relative} age estimates would not
 be significantly affected.
}.
However, the continuum map from our SINFONI data of BX\,482 has
insufficient S/N for our purposes, so we used the $H_{160}$ map as a proxy.
In what follows below, the H$\alpha$ equivalent widths are calculated in the
rest-frame (denoted $W^{\rm rest}({\rm H\alpha})$) assuming a flat $f_{\nu}$
continuum flux per unit frequency between rest-frame 5000\,\AA\ and 6563\,\AA\
and using the wavelength of H$\alpha$ for the conversion to continuum flux
per unit wavelength, $f_{\lambda}$.

We analyzed the NIC2 $H_{160}$ and SINFONI H$\alpha$ data of BX\,482 in
the same fashion as the NIC2 and ACS data of MD\,41, matching the NIC2
PSF to that of the SINFONI data, which has a $\rm FWHM = 0\farcs 17$.
The NIC2 and H$\alpha$ images were aligned relying primarily on the
outer isophotes and on the brightest clump visible in both maps (NIC2-\#1).
We refer the reader to \citeauthor*{FS11} for a full description of these
procedures.  The NIC2 $H_{160}$ band and SINFONI H$\alpha$ maps are shown
in Figure~\ref{fig-bx482_panel} along with a color-composite image created
from the PSF-matched maps.  Both the lower surface brightness component and
the brighter substructure are overall very comparable between the $H_{160}$
band and H$\alpha$ emission.  This similarity is reflected in the nearly
equal values of $G$, $\Psi$, and $M_{20}$ coefficients, and the structural
parameters $R_{\rm eff}$, $n$, and $b/a$ from both images agree well within
the $1\,\sigma$ uncertainties (see \citeauthor*{FS11}).

However, some differences do exist in the details of the spatial
distributions.  Specifically, clump \#4 identified in the NIC2 data
does not have an obvious counterpart in H$\alpha$, whereas there are
peaks in H$\alpha$, significantly offset to the north of clump \#4 and
on the northwestern side, that are not apparent in the NIC2 image.  We
thus ran the clump detection procedure on the H$\alpha$ map, finding
matches for all NIC2 clumps except \#4, and two additional clumps in
line emission, hereafter denoted ``H$\alpha$-\#1'' and ``H$\alpha$-\#2.''
The clumps are indicated on the maps of Figure~\ref{fig-bx482_panel}.
We measured the $H_{160}$ band and H$\alpha$ fluxes of each clump,
within the same aperture for consistent photometry, and both with
and without subtraction of the galaxy's background.  An aperture
correction was applied for the background-subtracted measurements. 
The same apertures as in \S~\ref{Sub-clumps_meth} were used for the
NIC2-identified clumps (the $H_{160}$ band magnitudes measured from
the original and PSF-matched NIC2 maps are not significantly different).
For the H$\alpha$-only clumps, the apertures were chosen based the
H$\alpha$ light profiles.  For clumps undetected in one or the other
map, we adopted $3\,\sigma$ upper limits on their background-subtracted
fluxes, derived from the noise properties of the images.
The flux measurements are given in Table~\ref{tab-clumpsprop_bx482}.

\subsubsection{Variations in H$\alpha$ Equivalent Width
               of the Clumps}
               \label{Sub-case_bx482_ewha}

Figure~\ref{fig-bx482_pixplots} shows in panel {\em (a)\/} the
distribution of $H_{160}$ band flux densities versus H$\alpha$ line
fluxes for individual pixels with $\rm S/N > 3$ in each map.  In panel
{\em (b)\/}, the results for the clumps (with and without background
subtraction) are plotted along with the galaxy-integrated measurements
and the fluxes summed over $\rm S/N > 3$ pixels for comparison\,\footnote{
 The small differences between the latter two is due to fewer pixels meeting
 the S/N criterion on the fainter northwest side than on the southeast side
 within the radius of $1\farcs 5$ of the total photometric aperture.
}.
Diagonal lines are overplotted in both panels, corresponding to
constant values of $W^{\rm rest}({\rm H\alpha})$ and labeled with
equivalent stellar ages for CSF models as described below.
Extinction does not have a significant effect on the
$W^{\rm rest}({\rm H\alpha})$ due to the proximity in wavelength
of H$\alpha$ and the rest-frame $\approx 5000$\,\AA\ region probed by
the $H_{160}$ bandpass; the effects of extinction run nearly parallel
to the lines of constant H$\alpha$ equivalent width in
Figure~\ref{fig-bx482_pixplots}.

Most clumps in BX\,482 contribute $\sim 1\% - 5\%$ of the total
emission in both $H_{160}$ band and H$\alpha$, and the clumps and
interclump regions span a comparable range of H$\alpha$ equivalent
widths.  While there are some differences for individual clumps
between the background-subtracted and raw flux estimates, the median
properties are almost the same.  In the ensemble, the clumps have
higher $W^{\rm rest}({\rm H\alpha})$ than the interclump regions,
with median values of 210\,\AA\ and 150\,\AA, respectively.
The two H$\alpha$-identified clumps have the highest
background-subtracted equivalent widths.
Two clumps are of particular interest in BX\,482: NIC2 clump \#1 and \#4.
Clump \#1 is the largest, most luminous, and one of the most massive of
all clumps identified in our NIC2 targets (see \S~\ref{Sub-clumps_res}).
It also has the largest contribution to the total host galaxy's emission,
$\sim 15\%$ in $H_{160}$ band as well as in H$\alpha$, roughly
$5 - 10\,\times$ higher than the average clump in our NIC2 targets.
However, clump \#1 does not differentiate in terms of its
$W^{\rm rest}({\rm H\alpha})$, which is equal to the median value
of all clumps in BX\,482.  In contrast, clump \#4 has about the same
size and fractional light contribution as the median values for the
clumps in our NIC2 targets, but it is one of the least massive clump.
It is further distinct among the BX\,482 clumps, with the lowest
measured $W^{\rm rest}({\rm H\alpha}) \sim 90 - 120$\,\AA,
depending on whether the galaxy background is subtracted or not.

\subsubsection{Variations in Ages of the Clumps and
               Implications for the $M/L$ Ratios}
               \label{Sub-case_bx482_impl}

To interpret the H$\alpha$ equivalent widths in terms of
stellar ages, we computed model predictions using the \citet{BC03}
evolutionary synthesis code with a \citet{Chab03} IMF and solar
metallicity.  The H$\alpha$ flux was calculated from the H ionizing
rates of the models, applying the recombination coefficients for case B
from \citet{Hum87} for an electron temperature $T_{\rm e} = 10^{4}~{\rm K}$
and density $n_{\rm e} = 10^{3}~{\rm cm^{-3}}$ \citep[see, e.g.,][]{FS09},
and the $H_{160}$-band flux density was approximated from the $g$-band
luminosity synthesized from the model spectra.  The effects of stellar
photospheric absorption on our H$\alpha$ flux measurements are negligible,
and so were not accounted for in the model predictions.  The variations
of $W^{\rm rest}({\rm H\alpha})$ with stellar age depend on the SFH.
The lines of constant $W^{\rm rest}({\rm H\alpha})$ plotted in
Figure~\ref{fig-bx482_pixplots} are labeled with the corresponding ages
for a CSF model.  As an example of alternative SFHs, for exponentially
declining SFRs with $e$-folding timescales $\tau$, the model curves for
the same ages in Figure~\ref{fig-bx482_pixplots} shift downwards along the
vertical axis and become more widely separated for ages $\ga \tau$.

For CSF, the $W^{\rm rest}({\rm H\alpha})$ measurements imply a median
stellar age of 55~Myr for the clumps and 140~Myr for the interclump
pixels.  For the large, luminous, and massive NIC2 clump \#1, its
$W^{\rm rest}({\rm H\alpha})$ also implies an age of 55~Myr, and so
it does not appear to stand out in its evolutionary stage compared to 
most other clumps.  The smaller, low-mass NIC2 clump \#4 with lowest
$W^{\rm rest}({\rm H\alpha})$ is the oldest clump in BX\,482, with
inferred age of $\rm \sim 240 - 570~Myr$ depending on the treatment
of the background galaxy light.  The impact of various SFHs, with SFR
declining or increasing with time, on the derived ages of the clumps
and interclump regions is illustrated in Appendix~\ref{App-WHa_models}.
In all cases, however, the relative ages are qualitatively unchanged as
long as the different regions have similar SFHs.  It is conceivable that
the duration of the star formation activity is shorter for the localized
kpc-sized clumps compared to the bulk of the stellar population across
the galaxy.  If so, most clumps would still be younger than the interclump
regions unless they follow exponentially increasing SFHs with timescales 
$\rm \tau \la 50~Myr$ (see Appendix~\ref{App-WHa_models}).

We can verify indirectly the validity of our hypothesis in
\S~\ref{Sub-clumps_meth} of constant observed $M/L$ for BX\,482,
on the basis of the relative ages and assuming that extinction plays
no role (i.e., that the relative dust obscuration is always the same
across the galaxy and between the clumps).  Using the same \citet{BC03}
models, and for a given SFH, the ages inferred for the clumps and the
interclump regions imply comparable $M/L$ ratios: the scatter among clumps
is $\rm \la 0.3~dex$, and the median $\log(M_{\star}/L_{g}^{\rm rest})$
values of the clumps are all within $\rm \approx 0.2~dex$ of each other
for different SFHs. 
These results for BX\,482 provide another example where the $M/L$ of the
clumps appears to be roughly constant and comparable to the ratio across the
entire galaxy, as we inferred from the analysis of the $i_{814} - H_{160}$
colors for MD\,41.  Obviously, this conclusion should not be generalized to
all galaxies, but it supports the plausibility of such an assumption in the
absence of better empirical constraints.

The analysis of the extinction-insensitive H$\alpha$ equivalent widths
in BX\,482 presented in this section revealed significant variations in
the relative stellar ages among the clumps.  If similar SFHs are applicable,
the data provide indications that all clumps have stellar populations of
comparable or younger ages than the interclump regions, with the exception
of one NIC2-identified clump without H$\alpha$ counterpart.
In addition, there is no evidence for variations by more than a factor
of a few in observed $M/L$ ratios between clumps and interclump regions.
The main uncertainty is in the SFH appropriate for different regions in
BX\,482, which cannot be constrained with the data currently available.
In future studies, detailed spatially-resolved constraints on the SFHs
across galaxies will help in this respect.  More accurate knowledge of
absolute ages will also be important to constrain clump lifetimes, a
crucial issue in the context of clump-driven scenarios for the evolution
of turbulent, gas-rich disks at high redshift
\citep[see the discussion by][and references therein]{Gen11}.

\section{DISCUSSION}
         \label{Sect-disc}

Our actively star-forming, rest-UV selected NIC2 targets exhibit clumps
in their rest-frame optical emission, just as clumps are observed in
the rest-frame UV emission of many $z \ga 1$ star-forming galaxies.
In the previous sections, we measured the clump properties and carefully
assessed the systematic uncertainties associated with their extraction,
in particular those related to the treatment of the host galaxy's
background light.
In this section, we now compare the properties of clumps in our NIC2
sample with those identified in other galaxy samples and at other
wavelengths.  We further explore constraints provided by our results
on the origin of clumps and their role in the evolution of the host
galaxies.

\subsection{Ensemble Properties of Clumps and their Origin}
            \label{Sub-clumps_ensemble}

\subsubsection{Comparison to Clumps in Other High-Redshift Galaxy Samples}
               \label{Sub-clumps_ensemble_obs}

The properties derived in \S~\ref{Sect-res} for the clumps in our
SINS NIC2 targets agree well with the ranges for clumps in high-redshift
galaxies derived from high resolution observations at other wavelengths.
The rest-frame optical sizes of $\rm \sim 1~kpc$ of our NIC2-identified
clumps are comparable to those measured in the rest-frame UV and H$\alpha$
line emission of clumps identified in {\em HST\/}/ACS optical imaging and
in AO-assisted SINFONI observations of $z \sim 1 - 3$ galaxies
\citep[e.g.,][]{Elm05a, Elm09a, Gen08, Gen11}.
The fractional contributions to the observed $H_{160}$ band emission in
our SINS NIC2 objects, as well as to the observed $i_{814}$ band emission
in MD\,41 and H$\alpha$ light in BX\,482, are typically a few percent per
clump (though with a wide range of $\sim 0.5\% - 16\%$), and together make
up $\sim 10\% - 25\%$ in each galaxy.
These fractions are also very similar to the fractions in ACS $i_{775}$ band
emission of rest-UV clumps in ``clump-clusters'' and ``chains,'' for which
\citet{Elm05a} and \citet{Elm09a} derive typically $\sim 2\%$ per clump and
a total of $\sim 25\%$ on average (ranging from $\la 1\%$ up to $\sim 50\%$)
for all clumps in a given galaxy.  Clump-clusters and chains are the most
prominently clumpy systems in optical imaging of high redshift galaxies in
the classification of \citet*{Elm04a}; rest-frame UV clumps in more regular
galaxies at similar redshifts tend to have lower fractional contributions
\citep{Elm09a}.  In H$\alpha$ emission, \citet{Gen11} determined comparable
fractions in the range $< 1\%$ up to $25\%$ for clumps in four $z \sim 2$
disks.  Thus, even if the clumps are a qualitatively prominent feature of
the morphologies, the apparently smoother and more extended emission
component generally dominates the total galaxy light.

In terms of stellar masses,
\citet{Elm05a} and \citet{Elm09a} found that clumps in clump-clusters
and chains appear to span a roughly constant range at $z \ga 0.5$ of
$M_{\star}^{\rm cl} \sim 10^{6} - 5 \times 10^{9}~{\rm M_{\odot}}$.
The clump stellar masses in our SINS NIC2 sample tend to lie in the top half
of the mass distributions of \citet{Elm09a}.  This may reflect higher host
stellar masses for most of our SINS NIC2 targets compared to the galaxies
studied by \citeauthor{Elm09a} \citep[see also the discussion by][]{Gen08}.
It may also possibly result from our clump selection in the rest-frame optical
instead of rest-frame UV as suggested by the analysis in \S~\ref{Sub-case_md41}
of MD\,41, where NIC2-identified clumps are overall more massive than
ACS-identified clumps.  Obviously, the important uncertainties involved
in estimating the clump properties, and the differences in methodology
--- notably in the treatment of the background light, and in the specifics
of the modeling used to estimate masses \citep[see][for details]{Elm09a} ---
mean that only order-of-magnitude comparisons can be made between the
different clump samples.  Nevertheless, these various studies indicate
that individual giant clumps typically represent each a few percent of
the total stellar mass of the host galaxies.

In the strongly-lensed $z \sim 1.5 - 3$ star-forming galaxies studied by
\citet{Jon10} and \citet[see also \citealt{Swi11}]{Swi10b}, the sizes and
dynamical masses of clumps identified in H$\alpha$ or in submillimeter dust
continuum and CO line emission tend to be lower (by factors of $\sim 2 - 10$)
than for the clumps in our SINS NIC2 objects.
Aided by linear magnification factors of $\sim 2 - 35$, the observations
for these lensed sources reach a source-plane spatial resolution down to
$\rm \approx 100~pc$ in the best cases, undoubtedly helping to resolve
smaller clumps.  However, the lensed galaxies are also on average roughly
twice smaller and an order of magnitude less massive than the SINS NIC2
galaxies, and have lower inferred $v_{\rm c}/\sigma_{0}$ ratios (implying
dynamically hotter disks).
The differences in clump properties between the SINS NIC2 and the lensed
samples may thus be more fundamentally related to the differences in host
galaxy properties, as expected in the theoretical framework of gas-rich
Toomre-unstable disks discussed next.

\subsubsection{Comparison to Theoretical Expectations for
               Fragmenting Gas-Rich, Turbulent Disks}
               \label{Sub-clumps_ensemble_theo}

Theoretical arguments as well as numerical simulations of turbulent
gas-rich Toomre-unstable disks indicate that they can fragment into
large and massive star-forming clumps
\citep[e.g.,][]{Nog99,Imm04a,Imm04b,Bour07,Gen08,Dek09,Age09,Cev10,Aum10,Genel10}.
All our disks have high inferred local intrinsic gas velocity dispersions
\citep[$\rm \sim 40 - 90~km\,s^{-1}$;][]{Gen08, Cre09} and appear to be
very gas-rich \citep[gas-to-baryonic mass fractions of $\sim 30\% - 70\%$;][]
 {Erb06, FS09, Tac10}.
Given these global properties, a possible interpretation of the clumps
seen in our five disks is that they are the result of disk instabilities.
Based on high S/N AO-assisted SINFONI observations, \citet{Gen11} argued
that giant clumps seen in H$\alpha$ and in rest-UV or optical broad-band
emission of several large $z \sim 2$ star-forming disks (including BX\,482)
generally correspond to highly unstable regions in the galaxies.  Indeed,
\citet{Gen11} estimated from the AO data Toomre $Q$ parameter values that
are well below unity at the clump locations.  Moreover, \citeauthor{Gen11}
also inferred $Q$ values in the interclump regions below unity, suggesting
global perturbations across the host disks.  These findings thus appear to
be consistent with the hypothesis that the observed clumps formed in-situ
from gravitational instabilities in gas-rich disks.

In this framework, the characteristic size and mass of the forming
clumps correspond to those of the fastest growing Jeans-unstable
fragmentation mode not stabilized by rotation, the Toomre scale length
and mass.  These characteristic scales are related to global properties
of the host galaxy and can be expressed as a function of disk radius and
mass, circular velocity $v_{\rm d}$, and local intrinsic gas velocity
dispersion $\sigma_{0}$ \citep[e.g.,][]{Gen08, Gen11, Esc08, Elm09b, Dek09}.
The ratio of $v_{\rm d}/\sigma_{0}$ is itself directly related to the disk
scale length and scale height, and is inversely proportional to the gas
mass fraction.  Applying Equations (5) of \citet{Gen11} with $Q \approx 1$
and the assumption of constant rotation velocity, the global galaxy parameters
for the five disks among our NIC2 sample \citep[from][]{Gen08, Cre09} imply
a typical Toomre radius of $\rm \sim 1~kpc$ and mass of
$\sim 4 \times 10^{9}~{\rm M_{\odot}}$.
Comparable values were obtained by \citet{Gen08} for a sample of
eight galaxies drawn from the SINS survey, with four objects in
common with our NIC2 sample (MD\,41, BX\,389, BX\,610, and BX\,482).
For $Q < 1$, these size and mass estimates would increase
(all other galaxy parameters being fixed).

The Toomre length and mass inferred above are of the same order as the
intrinsic sizes and stellar masses for the resolved clumps in the disks
of our NIC2 sample.  The total masses could be significantly larger than
the stellar masses if the clumps were themselves very gas-rich.  Assuming
that the typical gas-to-baryonic mass fraction of $\sim 50\%$ for the SINS
NIC2 disks is representative for the clumps, the inferred total mass for the
resolved disk clumps would be in even closer agreement with the expectations.
Direct molecular gas mass measurements for individual clumps in our targets
are not yet available.  However, estimates can be made for BX\,482 with
AO-assisted SINFONI data from H$\alpha$-based star formation rates and
using the ``Kennicutt-Schmidt'' relation.  Following \citet{Gen11},
to which we refer for details of the calculation and the choice of
Kennicutt-Schmidt relation (slope and zero point) that might be most
appropriate on the 1~kpc-scales of clumps in high-redshift star-forming
disks, the ``background-subtracted'' H$\alpha$ measurements imply
$M_{\rm gas}^{\rm cl}$ from
$\rm < 6 \times 10^{8}~M_{\odot}$ for clump \#2 to
$\rm \sim 5 \times 10^{9}~M_{\odot}$ for clump \#1
\citep[see also][their clump ``BX482-A'']{Gen11},
with a median of $\rm \sim 10^{9}~M_{\odot}$ (including limits).
These estimates include a $36\%$ correction for helium and metals, account
for the \citet{Chab03} IMF adopted in this paper, and assume that the global
$A_{V} = 0.8~{\rm mag}$ from the SED modeling of BX\,482 is representative of
the extinction towards the clumps.  The gas mass for clump \#1, the brightest
among our NIC2 sample, is comparable to those derived by \citet{Gen11} for
other bright clumps in three additional $z \sim 2$ disks.
For the NIC2-identified clumps in BX\,482, our results give
$M_{\rm gas}^{\rm cl}$ roughly twice the $M_{\star}^{\rm cl}$
estimates, and total (stellar $+$ gas) clump masses with median
$\rm \sim 1.5 \times 10^{9}~M_{\odot}$.  The implied gas-to-baryonic
mass fractions of $\sim 60 - 70\%$ for the clumps are comparable to
the galaxy-integrated value \citep{FS09}.

The theoretical expectations computed here as well as the empirical
measurements should be regarded as rough estimates in view of the
simplifying assumptions and the large uncertainties associated with
the determination of clump properties.  In addition, differences between
expectations and measurements can plausibly arise from different global
galaxy properties at the time when the observed stellar clumps were
formed and from evolution in clump properties (e.g., through mass loss).
Notwithstanding these limitations, the agreement between the predictions
and the observed properties is remarkable, and provides a framework to
interpret the prominent clumpy features in distant galaxies.
Similar conclusions were reached by \citet{Jon10} for the origin of clumps
in their lensed galaxy sample, based on the measured clump sizes and the
expectations computed with the size, mass, and dynamical properties of
the host disks.
As pointed out
in previous studies \citep[][and references therein]{Esc08, Elm09b, Gen11},
the high gas mass fractions and low $v_{\rm d}/\sigma_{0}$ ratios observed
in $z \ga 1$ disks naturally lead to the formation of larger self-gravitating,
star-forming complexes than in the less gas-rich, dynamically cold, and
geometrically thin $z \sim 0$ disks.  In addition, the presence of a
massive stabilizing stellar disk and/or bulge in local mature disk
galaxies also leads to smaller lower-mass star-forming complexes, or even
to the suppression of their formation \citep[e.g.,][]{Bour07, Dek09, Cev10}.

\subsection{Constraints on Radial Age Variations and Clump Evolution}
           \label{Sub-clumps_evol}

In the scenario outlined in the previous section,
the massive star-forming clumps are expected to migrate towards
the gravitational center as a result of dynamical friction against
the host disk and of clump-clump interactions.  Ultimately, the clumps
may coalesce into a young bulge within $\sim 10$ dynamical timescales
\citep{Nog99,Imm04a,Imm04b,Bour07,Gen08,Dek09,Age09,Cev10} unless they
are rapidly disrupted, for instance by stellar feedback or tidal torques.
In that respect, recent theoretical work has highlighted the possibly
dramatic impact on clump evolution of vigorous star formation feedback
\citep[][but see \citealt{Kru10} for a contrasting view]{Mur10, Genel10}.
High resolution and high S/N SINFONI observations have provided the first
empirical evidence of gas outflows originating from massive luminous
clumps in $z \sim 2$ disks \citep{Gen11}.  In some of the most actively
star-forming clumps, the outflow could be sufficiently strong to disperse
a large fraction of the initial gas before they reach the center of the
galaxies.  If the clumps do survive long enough and spiral inward, they
are then expected to exhibit a significant age spread, with older clumps
typically closer to the center of the galaxy \citep[e.g.,][]{Elm09a, Kru10}.

Our results for BX\,482 seem broadly consistent with this picture.
From measurements of the age-sensitive H$\alpha$ equivalent width, we
found in \S~\ref{Sub-case_bx482} indications of age variations among the
NIC2- and H$\alpha$-identified clumps.  Figure~\ref{fig-bx482_age_dproj}
shows the distribution of $W^{\rm rest}({\rm H\alpha})$ and of the
corresponding stellar ages assuming CSF as a function of deprojected
galactocentric distance $d$.  Although the data do not reveal a
significant correlation between $W^{\rm rest}({\rm H\alpha})$ or age
and $d$
(a Spearman rank correlation analysis yields a coefficient of
about 0.65 but with a significance of $\approx 1.5\,\sigma$), 
there is a clear distinction driven by the NIC2-identified
clump \#4 already highlighted in \S~\ref{Sub-case_bx482}
\footnote{
The distribution of $W^{\rm rest}({\rm H\alpha})$ with $d$ over the
interclump pixels in BX\,482 shows a large scatter with hardly any
trend ($\rho \approx 0.18$ at the $2\,\sigma$ level), reflecting the
overall very similar appearance of the H$\alpha$ and $H_{160}$
band emission discussed in \S~\ref{Sub-case_bx482}.}.
This clump, with
the lowest H$\alpha$ equivalent width and oldest age, is also nearest the
center of BX\,482 at $d = 2.9~{\rm kpc}$.  The six other clumps, with higher
equivalent widths and younger ages, are located at significantly larger
$d \approx 5 - 7.5~{\rm kpc}$.  This distinction is qualitatively
preserved for alternative choices of SFH as long as the clumps
follow similar SFHs (see Appendix~\ref{App-WHa_models}).

For MD\,41, the stellar ages cannot be constrained directly but the
deprojected radial distribution of $i_{814} - H_{160}$ colors of the
clumps can provide indications of maturity in terms of $M/L$ ratio.
The colors are plotted as a function of galactocentric distance $d$
in Figure~\ref{fig-md41_lmlg_dproj}, along with the derived
$M_{\star}/L_{g}^{\rm rest}$ ratios.  These plots suggest a weak trend
(with Spearman rank correlation coefficient of $\approx 0.85$ and
significance of $2\,\sigma$) of redder colors and
higher $M/L$ ratios at smaller radii; it is mostly dominated by clump
NIC2-\#1 and ACS-\#3, which, as noted in \S~\ref{Sub-case_md41_colvar},
exhibit the most distinct colors compared to the ensemble of clumps and
interclump regions
\footnote{
The distribution of $i_{814} - H_{160}$ colors with $d$ over the
interclump pixels in MD\,41 is almost flat, with increasing scatter
at larger radii (and $\rho \approx 0.15$ at the $2\,\sigma$ level),
as indicated by the fairly uniform colors across MD\,41 with the
exception of some of the clumps as highlighted in
\S~\ref{Sub-case_md41}.}.
If the dust extinction and the SFHs are similar
among the clumps, the radial distribution of colors and $M/L$ could
be indicative of older stellar ages for clumps at smaller galactocentric
distances.

In their studies of clumpy $z \sim 1 - 4$ galaxies in the H-UDF,
\citet[from ACS $B_{435}$, $V_{606}$, $i_{775}$, and $z_{850}$ band imaging]
 {Elm05a}
and
\citet[using the same ACS bands, complemented with NICMOS/NIC3
 $J_{110}$ and $H_{160}$ band imaging]
 {Elm09a}
found that ACS $i_{775}$ band-identified rest-frame UV clumps within
a given galaxy mostly span a narrow range of colors and that the
interclump regions tend to be redder, which they interpreted in terms of
older ages for the underlying stellar population across the host galaxy.
In $\sim 30\% - 50\%$ of the H-UDF chains and clump-clusters, and in all
their ``spiral'' types, \citet{Elm09a} also identified clumps that are
prominent in the near-IR bands, exhibit redder colors, and tend to lie
closer to the galaxies' centers.  These ``bulge-like clumps'' (or bulges
for spirals) have higher inferred stellar ages and masses than the other
clumps present within the same galaxy; they do not appear to differentiate
significantly in terms of star formation decay timescales or extinction.
Moreover, these differences in stellar ages and masses are more important
for spirals, less significant for clump-clusters and chains.  The derived
stellar ages of clumps span a wide range, centered around $\rm \sim 100~Myr$.
As argued by \citet[][see also \citealt{Elm09c}]{Elm09a}, the inferred
clump ages and the presence of a redder and more massive clump generally
closer to the center in clumpy and spiral types are consistent with inward
clump migration, and suggest that at least a fraction of the clumps may
survive long enough to reach the central few kpc.

The trends of redder colors for the clumps in MD\,41 and of older ages
for those in BX\,482 at smaller galactocentric radii are consistent with
the general findings of Elmegreen, Elmegreen, and coworkers.  Our higher
resolution NIC2 imaging, compared to NIC3, allowed us however to detect
clumps directly from longer wavelengths on $\sim 1~{\rm kpc}$ scales.
The NIC2 data furthermore provide a better match to the ACS PSF enabling
us, in the case of MD\,41, to identify and measure the properties of
clumps consistently in both the optical and near-IR bands.
While most ACS-identified clumps in MD\,41 tend to be bluer than the
interclump regions, in line with the results of \citet{Elm09a}, the
NIC2-identified clumps have in contrast comparable or redder colors.
The small but systematic differences in $i_{814} - H_{160}$ colors,
and in derived $M/L$ ratios and stellar masses, between the NIC2-
and ACS-identified clumps would suggest that clumps identified in the
rest-optical and rest-UV are complementary and probe together a wider
range in properties.  We can draw a similar conclusion for the clumps
identified in rest-frame optical and H$\alpha$ in BX\,482, with the
NIC2 and SINFONI PSFs also well matched to each other, although there
is more overlap here.  
The clump ages in BX\,482 for CSF models are compatible with the range
for clumps at $z \sim 2$ by \citet{Elm09a}; clump NIC2-\#4 closest to the
center lies at the high age end while all others are in the younger half
of the \citeauthor{Elm09a} distribution.

The dynamical timescale $t_{\rm dyn} = r / v$ at the effective
radius $\rm \approx 6~kpc$ of BX\,482 is $\rm \approx 25~Myr$
(correspondingly, the orbital timescale is
$t_{\rm orb}(r = R_{\rm eff}) \approx 160~{\rm Myr}$).
If the clumps have been forming stars at a roughly constant rate, the
stellar age of the oldest clump NIC2-\#4 thus implies it has survived
for $\sim 10$ to $\ga 20$ dynamical timescales, depending on whether
the raw or background-subtracted H$\alpha$ equivalent width estimates
is adopted.  It thus appears plausible that this clump has formed at
larger radii and has migrated to its present location closer to the
center at $d = 2.9~{\rm kpc}$.  The other younger clumps have ages
$\sim (1 - 4) \times t_{\rm dyn}$, so that they may still be located
fairly close to the radius at which they formed.  For the largest and
brightest clump NIC2-\#1, \citet{Gen11} detected a broad
($\rm FWHM \approx 400~km\,s^{-1}$) and somewhat blueshifted
(by several $\rm 10~km\,s^{-1}$) H$\alpha$ line component, which is
interpreted as a sign of outflowing gas.  Combining estimates of the
outflow kinematics and rate, and of the clump extent, gas mass, and
gas-phase oxygen abundance, \citet{Gen11} derived gas dispersion,
outflow expansion, and chemical enrichment timescales in the range
$\rm \sim 0.3 - 1~Gyr$.  These estimates would imply longer clump
lifetimes than our $W^{\rm rest}({\rm H\alpha})$-based stellar age
for CSF of $\rm \approx 55~Myr$, and may suggest that this clump
has survived for more than a couple dynamical timescales.

\subsection{Constraints on a Stellar Bulge/Inner Disk Component}
           \label{Sub-clumps_bulge}

Clump migration may provide an important mechanism for growing a
stellar bulge and/or inner disk component in high-redshift galaxies
\citep[e.g.][]{Nog99, Imm04a, Imm04b, Bour07, Car07, Elm09a, Cev10},
along with other internal mechanisms such as stellar bar-induced
streaming \citep[e.g.,][]{Bour02} or viscous drag in a gaseous disk
\citep[e.g.,][]{Lin87, Sil01}.  In $z \sim 0$ disks, these secular
processes occur on long timescales of at least several Gyrs
\citep[e.g.,][]{Kor04}.  At $z \sim 2$, however, disk galaxies are
significantly more gas-rich, and their gas phase appears to be much
more turbulent.  Under these conditions, secular processes should
proceed faster by an order of magnitude or more
\citep[e.g.][]{Nog99, Imm04a, Imm04b, Bour07, Gen08, Gen11, Cev10}.
If clumps are dispersed and lose mass before reaching the very center
of the galaxy, the build-up of a bulge component through clump migration
would be slower but this mechanism could still contribute to inward mass
transfer \citep[e.g.,][]{Bour07, Elm09c}.

From duty cycle arguments, \citet{Elm09a} inferred that the clumpy phase in
high-redshift disks may last for several Gyrs, or $5 - 10$ clump formation
epochs.  Some recent high-resolution cosmological simulations suggest that
rapid accretion from the halo along narrow, cold streams replenishing the
gas reservoir of high-redshift disks could help to maintain the clumpy
phase over such extended timescales \citep{Dek09, Cev10}.
One may therefore expect the presence of a more important stellar bulge
or inner disk component in more mature galaxies, as inward migration of
previous clump generations and secular processes would have had more
time to transport material in the central regions.

In this context, an immediate question is whether our data provide
evidence for a more important stellar bulge or inner disk component
in our NIC2 sample disks with older stellar populations.
At first glance, such a trend is not apparent from the observed spatial
distribution of the rest-frame optical continuum emission seen in our
NIC2 $H_{160}$-band images.  The best-fit S\'ersic index $n$ does not
show any significant correlation with best-fit stellar age from the
rest-UV/optical SED modeling, as discussed in \citeauthor*{FS11}.
The implied stellar mass fraction of the clump closest to the center
in each of the disks, if interpreted as a young bulge, also shows no
trend with galaxy stellar age.  More generally, none of the disks in
our sample shows obvious signs of a prominent centrally-concentrated
stellar component apart from BX\,663, for which the two-component
S\'ersic model fits in \citeauthor*{FS11} suggest that the central
peak could be associated with a fairly compact $n = 4$ bulge-like
component contributing $\approx 35\%$ of the total $H_{160}$-band
luminosity.  However, there could be an older stellar population in
the inner parts of the galaxies eluding detection even at rest-frame
$\approx 5000$\,\AA\ if it has a significantly higher $M/L$ ratio than
the stellar population dominating the observed rest-frame UV to optical SED.

To set constraints on the mass contribution of such a hidden stellar
component, we postulated that this component has an effective radius
$R_{\rm e} = 1~{\rm kpc}$ and S\'ersic index $n = 3$ akin to
the compact quiescent $z \sim 2$ $K$-selected galaxies of
\citet[][see also \citealt{Kri09}]{Dok08}, and assumed a fixed axis
ratio $b/a = 1$.  In what follows, we refer to this component as ``bulge,''
motivated by our choice of structural parameters, but we note that other
interpretations are not excluded.  We estimated how much extra light from
this component is allowed by the $1\,\sigma$ uncertainties of the observed
radial light profiles at $r < 2~{\rm kpc}$ for each of the five disks among
our NIC2 sample, as illustrated in Figure~\ref{fig-mass_core}.  To determine
the corresponding stellar mass contributions, we considered in turn different
$M_{\star}/L_{g}^{\rm rest}$ ratios.  Arguably, a realistic choice is the
typical $M/L$ found for $z \sim 2$ compact quiescent galaxies
\citep[see][]{Kri09}.  For the \citet{Dok08} sample, the median
$M_{\star}/L_{g}^{\rm rest}\,{\rm [M_{\odot}\,L_{g,\odot}^{-1}]} \approx 0.6$
based on the SED modeling of \citet{Kri08}, after adjustment to the
\citet{Chab03} IMF adopted in this work\,\footnote{
 The difference between $M_{\star}/L_{g}^{\rm rest}$ relevant for our
 data and $M_{\star}/L_{V}^{\rm rest}$ given by \citet{Kri08} is $< 10\%$
 for the typical ages, $A_{V}$, and SFHs of the quiescent galaxies.}.
This value is uncorrected for dust extinction, but the median $A_{V}$
for the quiescent sample is similar to that of our five SINS NIC2 disks
(0.7 versus 0.8~mag, respectively).  This ``passive'' $M/L$ is, however,
comparable to or lower than the ratio for our three disks with oldest
stellar ages.  If there was an important bulge characterized by such
a $M/L$ ratio, it would be clearly seen in the data.  We therefore
explored the more extreme scenarios in which all stars of the hidden
bulge are as old as the age of the universe at the redshift of each
disk, and whose light is either unobscured or attenuated by the same
amount as the stars that dominate the rest-UV/optical SED.  The $M/L$
for this maximally-old single stellar population (SSP) is calculated
from the same \citet{BC03} models for solar metallicity as used in
\citeauthor*{FS11} for the SED fitting.

The results are reported in Table~\ref{tab-mass_core}.  The light
contribution $\mathcal{F}^{\rm bulge}(H_{160})$ corresponds to the
fraction of the total observed $H_{160}$-band flux density for each
galaxy, and the mass fraction is relative to the stellar mass obtained
from the SED modeling,
$\mathcal{F}^{\rm bulge}(M_{\star}) \equiv 
 M_{\star}^{\rm bulge} / M_{\star}^{\rm SED}$.
The uncertainties given in Table~\ref{tab-mass_core} reflect the (small)
impact of varying the bulge S\'ersic index between $n = 2$ and 4, and of
increasing the $R_{\rm e}$ up to 3~kpc.  Unsurprisingly, the light fractions
are very small, in the range $1\% - 3.5\%$.  In terms of stellar mass, the
highest fractions hidden in an older bulge are found for the disks with
youngest ages (hence lowest $M_{\star}/L_{g}^{\rm rest}$ ratios) from the
best-fit SED model.  Nevertheless, the mass fractions are small to moderate
only, ranging from a few percent to $\approx 30\%$, with the exception of
MD\,41 (with estimates of up to a factor of $\sim 2.3$, but see below).
For all objects, the sum of $M_{\star}^{\rm SED} + M_{\star}^{\rm bulge}$
does not exceed the total dynamical mass \citep{Gen08, Cre09}, even when
accounting for the substantial amounts of gas inferred \citep{FS09, Tac10}.
A further plausibility check for our assumptions can be made for MD\,41
by comparing the resulting $M/L$ ratio for the various cases with the
$i_{814} - H_{160}$ color-based estimates in the inner $r \la 1~{\rm kpc}$.
Only the maximally-old obscured bulge scenario leads to a
$M_{\star}/L_{g}^{\rm rest}$ value that is largely different from the
ratio corresponding to the observed colors, suggesting that this case
(with more than twice the stellar mass being hidden) may be regarded
as unlikely.  More generally, the maximally-old SSP assumption is an
extreme one and the results should be considered as upper limits.

We conclude from the exercise above that the disks in our NIC2 sample
are unlikely to contain an important fraction of their stellar mass in
an old bulge or inner disk unseen in our $H_{160}$-band data, with the
possible exception of MD\,41.  BX\,663 appears to be the only object
in which a significant bulge-like component is directly visible, if
the AGN contribution to the rest-frame $\approx 5000$\,\AA\ emission
can be neglected.  The visible and hidden mass estimates add up to
$\sim 1 \times 10^{9} - 3 \times 10^{10}~{\rm M_{\odot}}$ in the
central $r \la 1~{\rm kpc}$ regions of our SINS NIC2 disks; thus,
it appears that none of these disks hosts a compact massive core.
No trend between stellar population age and the importance of a
central stellar component emerges when accounting for an old hidden
population.

Interestingly, \citet{Gen08} found in contrast evidence for a connection
between central dynamical mass concentration and maturity of the stellar
population, with the ratio of dynamical mass enclosed within $r \approx 3$
and 10~kpc increasing with older stellar age and higher inferred metallicity
of the galaxies.  For the four objects in common with our NIC2 sample,
\citeauthor{Gen08} derived 
$M_{\rm dyn}(r < {\rm 3\,kpc}) / M_{\rm dyn}(r < {\rm10\,kpc}) < 0.15$
for MD\,41, 0.21 for BX\,482, and 0.39 for both BX\,389 and BX\,610.
Combining the observed $H_{160}$-band light profiles from our NIC2 data
together with the constraints on the mass of a hidden evolved bulge, we
find that the implied stellar mass concentration calculated in the same
manner agrees reasonably well with the value inferred from the dynamics
for BX\,482 and BX\,610.  On the other hand, there are significant
differences between the central stellar and dynamical mass concentrations
for MD\,41 and BX\,389.

Thus, despite our findings in \S~\ref{Sect-cases} consistent with inward
clump migration --- a possible mechanism for growing stellar bulges or
inner disks over time --- our SINS NIC2 disks do not appear to show a
trend of higher central stellar mass concentration with older galaxy
age even when accounting for a possible hidden component.
Admittedly, our sample may be too small, and may sample too sparsely
different evolutionary stages between the objects with younger
($\rm \sim 50 - 320~Myr$) and older ($\rm \sim 2.5 - 2.8~Gyr$)
stellar ages, to discern such a trend.  Moreover,
various factors could obviously complicate the analysis presented in this
section, including the unknown spatial variations in extinction and in the
distribution of the gas, which contributes a large fraction of the total
mass in these galaxies.  Detailed maps constraining the extinction and gas
mass distribution, in addition to the $M/L$ ratio and the kinematics, will
be important to further test the scenario of bulge/inner disk growth through
clump migration and secular processes.

\subsection{Other Interpretations of the Clumpy Features}
            \label{Sub-clumps_interpr}

In this paper, we assumed that the compact regions of higher apparent
rest-frame optical surface brightness across our SINS NIC2 targets
represent genuine substructure in the distribution of the stellar
populations.  As we discussed earlier in this section, the properties
derived for these clumps as well as the global properties of the host
galaxies are consistent with the scenario in which giant massive clumps
result from gravitational instabilities in gas-rich, turbulent disks.
This scenario provides a framework to interpret clumpy morphologies
and to investigate dynamical processes that may play an important role
in the build-up of bulges and inner disks in high-redshift galaxies.

Other interpretations are possible.  Bright clumps could represent
merger components.  BX\,528 is a case where at least two of the clumps
identified most likely correspond to the progenitor galaxies of a major 
merging pair, and the southern clump \#-4 in BX\,389 is plausibly a small
companion galaxy (\citeauthor*[see][]{FS11}).  Clumps within disk galaxies
could also be of external origin, and correspond to newly accreted low-mass
systems.  The distinction between clumps formed in-situ and those accreted
from the halo is not easy.  Constraints could be obtained from an accurate
analysis of the stellar properties of clumps and host galaxies including
age and metallicity, or from a detailed baryonic and dynamical mass budget
of clumps to test whether they contain dark matter.
Arguably, it may seem however rather unlikely that as many as $5 - 10$
clumps per galaxy all originate from rapid accretion of satellites or
clumps brought in from the halo along filaments or dense, cold narrow
streams \citep[e.g.,][]{Elm08, Dek09, Age09, Cev10}.

Another possibility is that the kpc-sized clumps simply correspond to
locations across galaxies with lower line-of-sight dust obscuration.
Here again, the distinction between real substructure in the stellar
populations or star-forming regions and regions of lower obscuration
would necessitate detailed spectral or multi-wavelength constraints
on resolved scales of 1~kpc or smaller to pin-down accurately the
variations in extinction across the galaxies.
The role of extinction in clumpy morphologies is certainly a more
important concern in studies based on rest-frame UV imaging compared
to rest-frame optical imaging, but it could still play a significant
role at rest-frame $\approx 5000$\,\AA, as is well-known from nearby
dusty star-forming galaxies.  Small fractions of light of a few percent,
as inferred for most clumps in our SINS NIC2 sample as well as on average
for clumps in optical imaging of high-redshift galaxies, do not rule out
this interpretation, although it appears less plausible in the case of
very bright clumps with fractions in excess of $\sim 10\%$.
Extinction effects may also naturally explain the findings by
\citet{Elm09a} that rest-UV identified clumps tend to have bluer colors
than the interclump regions.  It may be more difficult to reconcile with
the trend of redder colors we found for rest-optically identified clumps
in MD\,41, or the ``bulge-like'' clumps often at more central locations
in galaxies discussed by \citet{Elm09a}.  If large localized spatial
variations in extinction are responsible for the clumpy features in our
galaxies, trends of redder colors and older ages at smaller radii, as
inferred for MD\,41 and BX\,482 in \S~\ref{Sub-case_bx482}, could reflect
age gradients in the underlying stellar population, dominated by younger
stars at larger radii and older ones in a bulge or inner disk component.

Possibly, clumpy features in high-redshift galaxies have multiple causes.
In future studies, high spatial resolution multiwavelength mapping and
sensitive spectroscopy will be essential to determine robustly the nature
of clumps.  More realistically in the near future, a combination of optical
and near-IR imaging (allowing spatially-resolved SED modeling), AO-assisted
integral field spectroscopy (mapping the kinematics and star formation,
and providing additional constraints on the age and SFH), and millimeter
interferometry (mapping directly the molecular gas distribution and kinematics)
will allow tighter constraints on the detailed physical, stellar, and dynamical
properties of clumps.  The first steps taken in this direction
\citep[e.g.,][]{Elm05a, Elm09c, Gen11, Tac10, Jon10, Swi10b, Swi11} show that,
if challenging, such an approach will provide key information necessary to
better understand the nature of the observed kpc-scale substructure and its
role in the evolution of distant galaxies.

\section{SUMMARY}   \label{Sect-conclu}

We have studied the characteristics of the clumpy features based on
{\em HST\/}/NIC2 $H_{160}$ band imaging for a unique sample of six
$z \sim 2$ star-forming galaxies with VLT/SINFONI integral field
spectroscopy.  The NIC2 imaging provides good sampling of the {\em HST\/}
PSF at 1.6\,\micron\ and allows us to detect reliably substructure in the
rest-frame optical emission on scales as small as $\rm 1.2~kpc$ at the
$z = 2.1 - 2.5$ of our targets.  We have identified ``clumps'' through a
systematic detection procedure that takes into account both host galaxy
background subtraction and PSF deconvolution for estimating the clump
luminosities, stellar masses, and sizes.  We assessed for the first time
in detail the systematic uncertainties associated with extracting clump
properties in distant galaxies, especially in terms of the treatment of
the underlying host galaxy light.  We interpreted the properties of the
clumps in the framework of gas-rich, turbulent disks in which giant
massive clumps can form through violent gravitational instabilities.
Our study complements previous work focussing on clumps identified
in the rest-frame UV or in H$\alpha$ line emission, and shows the
importance of including constraints at longer wavelengths and on
the same $\rm \sim 1~kpc$ scales for a more complete picture.

Our main results can be summarized as follows.

1. We identified between 2 and 7 clumps per galaxy in their rest-frame
optical emission.  Using the SINFONI data, we determined that these clumps
are unlikely to be caused predominantly by strong nebular emission lines
within the $H_{160}$ bandpass and thus must reflect compact sources of
stellar continuum light.
At least two of the clumps in the kinematically-identified binary merger
BX\,528 are likely associated with the merging galaxies.  For the large
kinematically-confirmed disk BX\,389, the clump most offset from the main
body of the galaxy can be securely identified as a small nearby companion
at a projected distance of $\rm \approx 5~kpc$.  Continuum light from a
stellar bulge component could plausibly dominate the central prominent
peak in BX\,663.
In terms of size, luminosity, and stellar mass, these specific clumpy
features do not differentiate significantly from the ensemble of all
clumps identified in our NIC2 targets, emphasizing the importance of
taking into account the nature of the host system from the dynamics,
morphologies, and spectral properties in the interpretation of ``clumps.''
All other clumps may have plausibly formed in-situ from fragmentation of
the turbulent, gas-rich disks of our targets.

2.
The stellar clumps in the $H_{160}$ band images of our NIC2 targets
generally represent surface brightness enhancements of $\sim 25\% - 35\%$
relative to the surrounding background from the host galaxy.
This contrast is substantially lower than for clumps identified in optical
ACS imaging of $z \sim 1 - 4$ prominently clumpy galaxies (clump-clusters
and chains), which generally appear to be brighter than the background by
factors of $\sim 2 - 4$ \citep[e.g.,][]{Elm09a}.
The contribution of individual clumps to the total galaxy $H_{160}$ band
emission is of a few percent typically, although with wide range from the
faintest ($\approx 0.5\%$) to the brightest ($\approx 16\%$).  The total
contribution of all clumps in a given galaxy is between $10\%$ and $25\%$.
The typical intrinsic rest-frame optical FWHM size of our clumps is
$\rm \sim 1~kpc$, similar to sizes of clumps identified and measured
in the rest-UV continuum and H$\alpha$ line emission
\citep[e.g.,][]{Elm05a, Elm09a, Gen08, Gen11}.
The inferred stellar masses have a median $\rm \sim 10^{9}~M_{\odot}$, and
a range that lies within the more massive half of the distribution inferred
for rest-UV identified clumps \citep{Elm05a, Elm09a}, possibly reflecting
the higher stellar masses of our galaxies and the clump identification
at longer wavelengths.

3.  
By combining the NIC2 data with {\em HST} ACS $i_{814}$ band imaging
available for MD\,41, and existing AO-assisted SINFONI H$\alpha$ data for
BX\,482, we inferred modest color, $M/L$, and stellar age variations among
most clumps within each galaxy, and between clumps and interclump regions.
In these two objects, the sets of clumps identified at different wavelengths
do not fully overlap.  In MD\,41, NIC2-identified clumps tend to have redder
$i_{814} - H_{160}$ colors than ACS-identified clumps and than the interclump
regions.  The colors for the reddest and bluest clumps appear to require
differences in stellar age in addition to extinction effects.
In BX\,482, we derived more robust constraints on the relative stellar
ages from the H$\alpha$ equivalent width, taken as the H$\alpha$ line
to rest-frame $\approx 5000$\,\AA\ continuum ratio.
Assuming similar star formation histories, the NIC2-identified clumps
tend to be older than the H$\alpha$-identified clumps but, except for
one, have ages younger than the interclump regions.
These findings suggest that clumps identified at different wavelengths
are complementary and probe together a wider range in clump properties.
Most importantly, clumps in MD\,41 and BX\,482 that are closer to the
center appear to show more maturity in terms of stellar populations.

4. The quantitative constraints on clump properties for the
kinematically-confirmed disks in our SINS NIC2 sample are consistent with
the scenario in which giant massive clumps form from disk instabilities.
Their sizes and masses are of the order of the Toomre scale length and mass
expected for disks with the same global properties as measured for the
galaxies (large gas velocity dispersion, high gas mass fractions, and
disk sizes).  
The trends of older and/or more obscured stellar populations of clumps
at smaller galactocentric radii for BX\,482 and MD\,41 are consistent
with inward migration of the massive clumps through dynamical friction
and clump-clump interactions.  These findings suggest that a fraction
of the clumps can survive to destruction by stellar feedback or other
disruptive mechanisms for at least a few dynamical timescales.
Our results add to other evidence from previous studies of
high-redshift clumpy galaxies \citep[e.g.,][]{Gen08, Elm09a}
and from numerical simulations of gas-rich unstable disks
\citep[e.g.,][]{Nog99, Imm04a, Imm04b, Bour07, Age09, Cev10} that
clump migration could be an important mechanism by which material
is transported in the inner regions of young disk galaxies,
contributing to the build-up of a bulge and inner disk component.

5.  We estimated for the disks in our sample
$\rm \sim 1 \times 10^{9} - 3 \times 10^{10}~M_{\odot}$
in a stellar bulge or inner disk component at $r \la 1 - 3~{\rm kpc}$,
with no obvious trend emerging among these five sources between central
stellar mass fraction and stellar age of the galaxy.  Our galaxies appear
to have roughly at least an order of magnitude lower {\em stellar} central
densities than the compact quiescent objects and even some of the star-forming
galaxies with significant putative compact cores among the massive $2 < z < 2.6$
NIC2 samples of \citet{Dok08} and \citet{Kri09}.
This could suggest different paths for the formation of the inner stellar
components \citep[see, e.g.,][and references therein]{Tac08, Fra08},
with highly dissipative mechanisms (such as major mergers) being more
important in the latter samples and secular processes dominating for
our SINS NIC2 sample.

The sample presented in this work is small and spans a modest range
in galaxy parameters, but our analysis highlights some of the key
issues that will be important to address in future studies.
The identification of clumps at different wavelengths can potentially
probe a wider range in clump properties and evolutionary stage.
Larger samples probing a wider range in host galaxy
properties and including (major) mergers as well will clearly allow
to better explore trends of clump properties with galaxy parameters,
evolutionary stage, and location within the host galaxy, and to assess
possible differences between clumps in disks and in merging systems.
The combination of $M/L$-sensitive and age-sensitive properties will be
crucial to make progress.
One of the currently most important limitations in the interpretation
of the clump properties is their unknown star formation history.
Tighter constraints on clump masses, star formation and outflow rates,
and internal structure and dynamics will provide better insights into
the issue of clump lifetimes and their role in the internal dynamical
evolution of young galaxies.

\acknowledgments

We are grateful to our many colleagues for stimulating discussions
and insightful comments on various aspects of this work, in particular
A. Burkert, P. Johansson, T. Naab, A. Renzini, S. Wuyts, and the entire
SINS team.
We also wish to thank M. Swinbank and T. Jones for useful discussions
and additional information about clumps in their lensed galaxy samples.
We thank the referee for a thoughtful report and useful suggestions
that improved the presentation of the results.
Support for HST program \# 10924 was provided by NASA through a grant
from the Space Telescope Science Institute, which is operated by the
Association of Universities for Research in Astronomy, Inc., under
NASA contract NAS 5-26555.
N.M.F.S. acknowledges support by the Minerva program of the MPG.
A.E.S. acknowledges support from the David and Lucile Packard
Foundation.
N.B. is supported by the Marie Curie grant PIOF-GA-2009-236012 from the
European Commission.
G.C. acknowledges support by ASI-INAF grant I/009/10/0.
A.S. thanks the DFG for support via German-Israeli Project
Cooperation Grant STE1868/1-1.GE625/15.1.

\vspace{2.0cm}

\appendix

\section{VARIATIONS OF CLUMP STELLAR AGES WITH DIFFERENT
         STAR FORMATION HISTORIES FOR BX\,482}
         \label{App-WHa_models}

In \S\S~\ref{Sub-case_bx482} and \ref{Sub-clumps_evol}, we used
the H$\alpha$ equivalent width, with the continuum flux density
approximated from the $H_{160}$-band photometry of our targets,
to derive stellar ages for the the clumps and interclump regions
in BX\,482.  This indicator is nearly insensitive to interstellar
dust reddening if differential extinction is negligible, i.e., if
the H$\alpha$ line emission from the \ion{H}{2} regions is attenuated
by the same amount as the continuum light from the stellar population.
On the other hand, the $W^{\rm rest}({\rm H\alpha})$ depends sensitively
on the star formation history (SFH).  The analysis throughout the paper
relies on the simple assumption of constant star formation (CSF) with age.
Here we explore the effects of adopting other SFHs on the interpretation
of the $W^{\rm rest}({\rm H\alpha})$.

For this purpose, we used the same \citet{BC03} models of solar metallicity,
with the \citet{Chab03} IMF and the \citet{Cal00} reddening law as for the
SED modeling of the integrated rest-UV/optical SED of our SINS NIC2 sample
galaxies.  We investigated three different generic forms for the SFH:
(1) exponentially declining models with
    ${\rm SFR}(t) \propto \exp(-t / \tau)$
    (``$\tau$'' models),
(2) exponentially increasing models with
    ${\rm SFR}(t) \propto \exp(t / \tau)$
    (``inverted $\tau$'', or ``i$\tau$'' models), and
(3) a hybrid case with
    ${\rm SFR}(t) \propto (t / \tau)\,\exp(-t / \tau)$
    (``delayed $\tau$'', or ``d$\tau$'' models),
where {\rm SFR} is the star formation rate, $t$ is the time elapsed since
the onset of star formation (taken as the age of the stellar population),
and $\tau$ is the characteristic timescale of the star formation.
A CSF corresponds, for the exponentially-declining parametrization,
to $\tau \rightarrow \infty$.
We computed suites of models for a range of $\tau$ from 10~Myr to 1~Gyr.
For all SFHs, the $W^{\rm rest}({\rm H\alpha})$ decreases monotonically
with age; differences are in the shape of the curves.

We derived the stellar ages and uncertainties of the clumps in BX\,482
from the $W^{\rm rest}({\rm H\alpha})$ and their $1\,\sigma$ errors
(projected onto the model curves).  We always assumed a fixed maximum
age possible corresponding to the age of the Universe at the redshift
of BX\,482.  Ages older than the Universe are formally obtained only for
NIC2 clump \#4, with lowest $W^{\rm rest}({\rm H\alpha})$, and inverted
$\tau$ models with timescales $\rm \ga 100~Myr$.  This suggests that either
such SFHs are inappropriate or there is significant differential attenuation
between the H$\alpha$ line and stellar continuum emission; the data currently
available do not allow us to distinguish between these two possibilities.
Figure~\ref{fig-WHa_models} illustrates the differences in ages obtained
for each SFH family and three representative timescales, $\tau = 30$, 100,
and 300 Myr, plotted as a function of the ages for CSF models.
Because of the monotonic variations with $t$ irrespective of the SFH,
the relative age differences between clumps are qualitatively preserved
for any given SFH.  The range of ages tends to become narrower for shorter
star formation timescales.  However, the uncertainties of the measurements
are typically sufficiently small that the age variations among the clumps
remain significant for all $\tau$ values considered.
This result holds independently of the treatement of the background light
from the host galaxy; the measurements based on the ``background-subtracted''
photometry lead to larger age differences than those based on the ``raw''
photometry, i.e., not accounting for the contribution of the underlying
host stellar population to the clump light.

From this exploration of the impact of SFHs on the interpretation of
the $W^{\rm rest}({\rm H\alpha})$, we conclude the following.  While the
absolute ages of the clumps in BX\,482 and ranges thereof depend on the
SFH, the relative age variations are qualitatively robust and quantitatively
significant in view of the measurements uncertainties.  This conclusion is
valid as long as the SFHs of individual clumps are similar in their general
time dependence and in their timescale.  Verifying this assumption would be
important but obviously, detailed constraints on clump SFHs would require
much more extensive multiwavelength data sets at high angular resolution.
Future studies with HST/WFC3 and later with JWST and the ELTs should
be able to provide such constraints and further insights in the
detailed star formation history across individual galaxies, thereby
also allowing more accurate reconstruction of the galaxy-wide global
SFH, one of the major uncertainty in SED modeling of distant objects.


\clearpage


\setcounter{figure}{0}
\setcounter{section}{0}

\tabletypesize{\small}
\begin{deluxetable}{lllllll}
\tablecolumns{7}
\tablewidth{0pt}
\tablecaption{Galaxies observed
              \label{tab-targets}}
\tablehead{
   \colhead{Property} & 
   \colhead{Q1623-BX528} & 
   \colhead{Q1623-BX663\,\tablenotemark{a}} & 
   \colhead{SSA22a-MD41} & 
   \colhead{Q2343-BX389} & 
   \colhead{Q2343-BX610\,\tablenotemark{a}} & 
   \colhead{Q2346-BX482}
}
\startdata
$z_{\rm H\alpha}$\,\tablenotemark{b}        &
2.2683 & 2.4332 & 2.1704 & 2.1733 & 2.2103 & 2.2571 \\[0.5ex]
$H_{\rm 160}$~(mag)\,\tablenotemark{c}      &
$22.33 \pm 0.06$ & $22.79 \pm 0.10$ & $22.64 \pm 0.05$ &
$23.11 \pm 0.10$ & $22.09 \pm 0.06$ & $22.34 \pm 0.07$ \\[0.5ex]
Age~(Myr)\,\tablenotemark{d}                &
$2750^{+96}_{-2110}$   & $2500^{+147}_{-800}$  &  $50^{+31}_{-0}$    &
$2750^{+224}_{-1945}$  & $2750^{+173}_{-650}$  & $321^{+485}_{-141}$ \\[0.5ex]
$A_{V}$~(mag)\,\tablenotemark{d}            &
$0.6 \pm 0.2$  & $0.8 \pm 0.2$  & $1.2 \pm 0.2$  &
$1.0 \pm 0.2$  & $0.8 \pm 0.2$  & $0.8 \pm 0.2$  \\[0.5ex]
$M_{\star}$~($\rm 10^{10}~M_{\odot}$)\,\tablenotemark{d}     &
$6.95^{+0.17}_{-3.61}$  & $6.40^{+0.22}_{-2.28}$  &
$0.77^{+0.10}_{-0.03}$  & $4.12^{+0.77}_{-2.16}$  &
$10.0^{+2.7}_{-0.6}$    & $1.84^{+0.79}_{-0.46}$  \\[0.5ex]
$M_{\star}/L_{g}^{\rm rest}$~($\rm M_{\odot}\,L_{g,\odot}^{-1}$)\,\tablenotemark{d} &
0.51  & 0.58  & 0.08  & 0.78  & 0.63  & 0.15  \\[0.5ex]
SFR~($\rm M_{\odot}\,yr^{-1}$)\,\tablenotemark{d}            &
$42^{+29}_{-16}$  & $42^{+13}_{-12}$  & $185^{+3}_{-70}$  &
$25^{+17}_{-2}$   & $60^{+26}_{-1}$   & $80^{+42}_{-32}$  \\[0.5ex]
sSFR~($\rm Gyr^{-1}$)\,\tablenotemark{d}                     &
$0.6^{+1.7}_{-0.1}$   & $0.7^{+0.3}_{-0.2}$   & $24^{+1}_{-9}$        &
$0.6^{+1.2}_{-0.1}$   & $0.6 \pm 0.2$         & $4.3^{+3.0}_{-2.5}$   \\[0.5ex]
Kinematic Type\,\tablenotemark{e}           &
Major merger  & Disk  & Disk  & Disk  & Disk  & Disk  \\[0.5ex]
$R_{\rm e}$~(kpc)\,\tablenotemark{f}        &
$4.86^{+0.13}_{-0.10}$  & $4.54^{+7.71}_{-0.79}$  &
$5.69^{+0.20}_{-0.13}$  & $5.93^{+0.17}_{-0.12}$  &
$4.44 \pm 0.08$         & $6.22^{+0.13}_{-0.12}$  \\[0.5ex]
$n$\,\tablenotemark{f}                      &
$0.16 \pm 0.01$         & $2.00^{+2.11}_{-0.59}$  &
$0.54^{+0.07}_{-0.06}$  & $0.36^{+0.07}_{-0.05}$  &
$0.57 \pm 0.04$         & $0.14 \pm 0.01$         \\[0.5ex]
$b/a$\,\tablenotemark{f}                    &
$0.31 \pm 0.01$      & $0.73^{+0.04}_{-0.02}$  & $0.31 \pm 0.01$      &
$0.30 \pm 0.01$      & $0.56 \pm 0.01$         & $0.48 \pm 0.01$     \\[0.5ex]
\enddata
\tablenotetext{a}
{
Spectral signatures of a Type~2 AGN are detected in the optical and near-IR
spectrum of BX\,663 and the large-scale H$\alpha$ kinematics suggest the
host galaxy is a low-inclination rotating disk \citep{Erb06, FS06, FS09}.
The {\em Spitzer\/}/MIPS 24\,\micron\ flux in BX\,663 is consistent with
obscured AGN activity.  A 24\,\micron\ excess is also detected for BX\,610
but since there is no other signature of AGN activity in this galaxy, it
could be due to enhanced mid-IR PAH emission around rest-frame
8\,\micron\ instead (see \S~\ref{Sub-sampleprop}).
}
\tablenotetext{b}
{
Systemic vacuum redshift derived from the integrated H$\alpha$ line
emission \citep{FS09}.
}
\tablenotetext{c}
{
The $H_{160}$ band magnitudes (AB system) measured from the NIC2
data in a circular aperture with diameter of $3^{\prime\prime}$.
}
\tablenotetext{d}
{
Stellar age, visual extinction ($A_{V}$), stellar mass ($M_{\star}$),
stellar mass to rest-frame SDSS $g$-band luminosity ratio
($M_{\star}/L_{g}^{\rm rest}$, uncorrected for extinction), and
absolute and specific star formation rate (SFR and sSFR) of the best-fit
evolutionary synthesis model to the observed optical to near-IR photometry
of the galaxies.  The SEDs were fitted using \citealt{BC03} models with solar
metallicity, a \citealt{Chab03} IMF, the \citealt{Cal00} reddening law, and a
constant SFR.
The formal (random) fitting uncertainties listed are derived from
the 68\% confidence intervals based on 200 Monte Carlo simulations;
systematic uncertainties (from SED modeling assumptions) are estimated
to be typically a factor of 1.5 for the stellar masses, $\rm \pm 0.3~mag$
for the extinctions, and factors of $\sim 2 - 3$ for the ages as well
as for the absolute and specific star formation rates.
}
\tablenotetext{e}
{
Classification according to the H$\alpha$ kinematics from SINFONI
\citep{Sha08, FS06, FS09}.
}
\tablenotetext{f}
{
Effective radius, light concentration index, and ratio of minor
to major axis ($R_{\rm e}$, $n$, $b/a$) obtained from single-component
S\'ersic model fits to the two-dimensional $H_{160}$ band surface brightness
distributions of the galaxies; the best-fit parameters correspond to the
median of the distribution of results from 500 GALFIT runs as described
in \citeauthor*{FS11}, and the uncertainties represent the 68\% confidence
intervals around the best-fit value.
For BX\,528, the two main merging components ``SE'' and ``NW'' are resolved
in the NIC2 data; the results of simultaneous two-component S\'ersic model
fits are 
$R_{\rm e} = 3.18^{+0.18}_{-0.25}$,
$n = 1.20^{+0.15}_{-0.19}$, and $b/a = 0.35^{-0.08}_{-0.03}$
for BX\,528$-$SE, and
$R_{\rm e} = 3.57^{+1.98}_{-0.57}$,
$n = 2.91^{+1.70}_{-1.34}$, and $b/a = 0.51^{+0.06}_{-0.05}$
for BX\,528$-$NW.
For BX\,389, the NIC2 data resolve the small companion south of the main
part of the galaxy; the results of simultaneous two-component S\'ersic
model fits are
$R_{\rm e} = 5.89^{+0.08}_{-0.09}$,
$n = 0.32^{+0.04}_{-0.04}$, and $b/a = 0.27 \pm 0.01$
for the main body of BX\,389, and
$R_{\rm e} = 2.71^{+2.18}_{-0.69}$,
$n = 3.03^{+2.64}_{-1.75}$, and $b/a = 0.20^{+0.35}_{-0.16}$
for BX\,389$-$South.
}
\end{deluxetable}

\clearpage

\tabletypesize{\footnotesize}
\begin{deluxetable}{ccccccccccccc}
\tablecolumns{13}
\setlength{\tabcolsep}{0.10cm}
\tablewidth{0pt}
\tablecaption{Properties of clumps identified in the NIC2 images
              \label{tab-clumpsprop}}
\tablehead{
   \colhead{Clump\,\tablenotemark{a}} &
   \colhead{$d_{\rm proj}$\,\tablenotemark{b}} &
   \colhead{$d$\,\tablenotemark{b}} &
   \colhead{$r_{\rm phot}$\,\tablenotemark{c}} &
   \colhead{Ap. corr.\,\tablenotemark{c}} &
   \colhead{FWHM~\tablenotemark{d}} &
   \colhead{$\rm FWHM^{0}$~\tablenotemark{d}} &
   \colhead{$\mathcal{F}^{\rm cl}_{\rm PS}$~\tablenotemark{e}} &
   \colhead{$\mathcal{F}^{\rm cl}_{\rm raw}$~\tablenotemark{e}} &
   \colhead{$\mathcal{F}^{\rm cl}$~\tablenotemark{e}} &
   \colhead{$H_{\rm 160}^{\rm cl}$~\tablenotemark{f}} &
   \colhead{$L_{g}^{\rm rest,cl}$~\tablenotemark{g}} &
   \colhead{$M_{\star}^{\rm cl}$~\tablenotemark{h}} \\
   \colhead{} &
   \colhead{(kpc)} &
   \colhead{(kpc)} &
   \colhead{(arcsec)} &
   \colhead{} &
   \colhead{} &
   \colhead{(kpc)} &
   \colhead{} &
   \colhead{} &
   \colhead{} &
   \colhead{(mag)} &
   \colhead{(${\rm L}_{g,\,\odot}$)} &
   \colhead{($\rm M_{\odot}$)}
}
\startdata
\cutinhead{$\rm Q1623-BX528$}
  1  &  3.92   &  $4.0 \pm 0.3$  &  0.125  &  1.79  &   1.23   &  0.31  &
        0.016  &  0.093  &  0.052  &
        25.53  &  $6.5 \times 10^{9}$  &  $3.6 \times 10^{9}$ \\
  2  &  2.52   &  $2.6 \pm 0.3$  &  0.125  &  1.79  &   1.52   &  0.94  &
        0.011  &  0.092  &  0.040  &
        25.82  &  $5.0 \times 10^{9}$  &  $2.8 \times 10^{9}$ \\
  3  &  0.86   &  $2.6 \pm 0.5$  &  0.125  &  1.79  &   1.27   &  0.43  &
        0.008  &  0.054  &  0.023  &
        26.42  &  $2.9 \times 10^{9}$  &  $1.6 \times 10^{9}$ \\
  4  &  5.02   &  $5.2 \pm 0.6$  &  0.200  &  1.44  &   1.35   &  0.64  &
        0.022  &  0.108  &  0.077  &
        25.11  &  $9.7 \times 10^{9}$  &  $5.4 \times 10^{9}$ \\
All  & \ldots  & \ldots  & \ldots  & \ldots  & \ldots  & \ldots &
        0.057  &  0.347  &  0.193  &
       \ldots  &  \ldots  &  \ldots  \\
\cutinhead{$\rm Q1623-BX\,663$}
  1  &  2.56   &  $2.7 \pm 0.7$  &  0.100  &  2.09  &  $<1.18$ & \ldots &
        0.008  &  0.035  &  0.012  &
        27.62  &  $1.1 \times 10^{9}$  &  $7.5 \times 10^{8}$ \\
  2  &  0.68   &  $0.68 \pm 0.08$  &  0.225  &  1.33  &   1.81   &  1.37  &
        0.022  &  0.209  &  0.132  &
        24.99  &  $1.2 \times 10^{10}$ &  $8.4 \times 10^{9}$ \\
All  & \ldots  & \ldots  & \ldots  & \ldots & \ldots   & \ldots &
        0.030  &  0.244  &  0.143  &
       \ldots  &  \ldots  &  \ldots  \\
\cutinhead{$\rm SSA22a-MD\,41$}
  1  &  0.56   &  $1.1 \pm 0.2$  &  0.100  &  2.09  &  $<1.20$ & \ldots &
        0.008  &  0.038  &  0.017  &
        27.08  &  $1.5 \times 10^{9}$  &  $1.3 \times 10^{8}$ \\
  2  &  1.65   &  $2.3 \pm 0.2$  &  0.100  &  2.09  &  $<1.20$ & \ldots &
        0.009  &  0.044  &  0.021  &
        26.82  &  $1.9 \times 10^{9}$  &  $1.6 \times 10^{8}$ \\
  3  &  2.96   &  $3.2 \pm 0.1$  &  0.100  &  2.09  &  $<1.20$ & \ldots &
        0.006  &  0.039  &  0.012  &
        27.45  &  $1.0 \times 10^{9}$  &  $9.2 \times 10^{7}$ \\
  4  &  4.77   &  $5.8 \pm 0.4$  &  0.100  &  2.09  &  $<1.20$ & \ldots &
        0.011  &  0.036  &  0.026  &
        26.58  &  $2.3 \times 10^{9}$  &  $2.0 \times 10^{8}$ \\
All  & \ldots  & \ldots  & \ldots  & \ldots & \ldots   & \ldots &
        0.035  &  0.156  &  0.076  &
       \ldots  &  \ldots  &  \ldots  \\
\cutinhead{$\rm Q2343-BX\,389$}
  1  &  5.65   &  $5.7 \pm 0.1$  &  0.175  &  1.53  &   1.74   &  1.26  &
        0.009  &  0.077  &  0.043  &
        26.53  &  $2.4 \times 10^{9}$  &  $1.8 \times 10^{9}$ \\
  2  &  2.57   &  $3.3 \pm 0.4$  &  0.150  &  1.63  &   1.72   &  1.23  &
        0.008  &  0.070  &  0.030  &
        26.92  &  $1.7 \times 10^{9}$  &  $1.2 \times 10^{9}$ \\
  3  &  0.67   &  $0.77 \pm 0.06$  &  0.125  &  1.79  &  $<1.20$ & \ldots &
        0.009  &  0.054  &  0.021  &
        27.29  &  $1.2 \times 10^{9}$  &  $8.8 \times 10^{8}$ \\
  4  &  5.10   &  \ldots         &  0.175  &  1.53  &   1.38   &  0.69  &
        0.012  &  0.046  &  0.032  &
        26.83  &  $1.8 \times 10^{9}$  &  $1.3 \times 10^{9}$ \\
  5  &  3.78   &  $4.4 \pm 0.4$  &  0.200  &  1.44  &   1.63   &  1.10  &
        0.021  &  0.134  &  0.079  &
        25.87  &  $4.4 \times 10^{9}$  &  $3.2 \times 10^{9}$ \\
  6  &  5.64   &  $5.9 \pm 0.3$  &  0.150  &  1.63  &   1.24   &  0.32  &
        0.010  &  0.052  &  0.017  &
        27.53  &  $9.6 \times 10^{8}$  &  $7.0 \times 10^{8}$ \\
All  & \ldots  & \ldots  & \ldots  & \ldots & \ldots   & \ldots &
        0.068  &  0.433  &  0.222  &
       \ldots  &  \ldots  &  \ldots  \\
\cutinhead{$\rm Q2343-BX\,610$} 
  1  &  4.11   &  $4.3 \pm 0.2$  &  0.125  &  1.79  &   1.57   &  1.02  &
        0.006  &  0.045  &  0.020  &
        26.32  &  $3.0 \times 10^{9}$  &  $2.0 \times 10^{9}$ \\
  2  &  0.62   &  $0.63 \pm 0.02$  &  0.175  &  1.53  &   1.48   &  0.87  &
        0.015  &  0.107  &  0.039  &
        25.60  &  $5.9 \times 10^{9}$  &  $3.9 \times 10^{9}$ \\
  3  &  3.16   &  $3.5 \pm 0.2$  &  0.175  &  1.53  &   1.51   &  0.92  &
        0.009  &  0.096  &  0.020  &
        26.34  &  $3.0 \times 10^{9}$  &  $2.0 \times 10^{9}$ \\
  4  &  1.44   &  $1.5 \pm 0.1$  &  0.125  &  1.79  &   1.43   &  0.77  &
        0.007  &  0.059  &  0.018  &
        26.45  &  $2.7 \times 10^{9}$  &  $1.8 \times 10^{9}$ \\
  5  &  4.91   &  $5.3 \pm 0.3$  &  0.150  &  1.63  &  $<1.20$ & \ldots &
        0.005  &  0.043  &  0.009  &
        27.23  &  $1.3 \times 10^{9}$  &  $8.8 \times 10^{8}$ \\
  6  &  1.59   &  $2.7 \pm 0.3$  &  0.150  &  1.63  &  $<1.20$ & \ldots &
        0.004  &  0.063  &  0.005  &
        27.93  &  $6.9 \times 10^{8}$  &  $4.6 \times 10^{8}$ \\
  7  &  3.22   &  $3.6 \pm 0.3$  &  0.150  &  1.63  &   1.20   &  0.12  &
        0.014  &  0.083  &  0.044  &
        25.48  &  $6.6 \times 10^{9}$  &  $4.4 \times 10^{9}$ \\
All  & \ldots  & \ldots  & \ldots  & \ldots & \ldots   & \ldots &
        0.060  &  0.496  &  0.155  &
       \ldots  &  \ldots  &  \ldots   \\
\cutinhead{$\rm Q2346-BX\,482$} 
  1  &  4.74   &  $4.8 \pm 0.2$  &  0.275  &  1.14  &   1.93   &  1.51  &
        0.015  &  0.247  &  0.158  &
        24.34  &  $1.9 \times 10^{10}$ &  $2.9 \times 10^{9}$ \\
  2  &  3.02   &  $5.6 \pm 0.9$  &  0.150  &  1.63  &  $<1.19$ & \ldots &
        0.005  &  0.035  &  0.017  &
        26.75  &  $2.1 \times 10^{9}$  &  $3.2 \times 10^{8}$ \\
  3  &  4.36   &  $6.7 \pm 1.3$  &  0.175  &  1.53  &   1.83   &  1.39  &
        0.007  &  0.060  &  0.018  &
        26.71  &  $2.2 \times 10^{9}$  &  $3.3 \times 10^{8}$ \\
  4  &  1.97   &  $2.9 \pm 0.6$  &  0.200  &  1.44  &   1.59   &  1.06  &
        0.007  &  0.081  &  0.027  &
        26.28  &  $3.3 \times 10^{9}$  &  $4.9 \times 10^{8}$ \\
  5  &  3.68   &  $7.6 \pm 1.0$  &  0.175  &  1.53  &   1.26   &  0.41  &
        0.010  &  0.050  &  0.030  &
        26.17  &  $3.6 \times 10^{9}$  &  $5.4 \times 10^{8}$ \\
All  & \ldots  & \ldots  & \ldots  & \ldots & \ldots   & \ldots &
        0.045  &  0.473  &  0.250  &
        \ldots  &  \ldots  &  \ldots 
\enddata
\tablenotetext{a}
{
Here, ``clump'' refers generically to $\geq 3\,\sigma$ peaks on
characteristic spatial scales of $1 - 1.5$ times the PSF FWHM above the
local background in the $H_{160}$ band images (\S~\ref{Sub-clumps_meth}).
Clumps \#1 and \#2 of the major merger BX\,528 are part of the
SE component while clump \#4 is associated with the NW component.
In BX\,663, clump \#2 is close to the center of the galaxy and may
be associated with a stellar bulge or the AGN present in this object.
Clump \#4 of BX\,389 corresponds to the small companion to the
south of the main galaxy.
}
\tablenotetext{b}
{
Projected distance ($d_{\rm proj}$) relative to the geometric center
of the galaxies, and deprojected distance ($d$) assuming a circular
configuration with inclination angle corresponding to the best-fit
axis ratio from the NIC2 morphology (Table~\ref{tab-targets}).
No projected distance is calculated for the southern companion in BX\,389
(clump \#4), which appears to be located below the galactic plane.
}
\tablenotetext{c}
{
Radius of the circular aperture used to measure the clump fluxes,
and aperture correction based on the curve-of-growth from the empirical
PSF light profile, corresponding to the flux ratio in circular apertures
of radius 1\arcsec\ and $r_{\rm phot}$.
}
\tablenotetext{d}
{
FWHM is the observed direct full-width at half-maximum from the clump
radial light profiles after subtraction of the local background.
$\rm FWHM^{0}$ is the intrinsic size derived by subtracting in
quadrature the observed FWHM and the PSF FWHM.  For unresolved
clumps, the PSF FWHM is taken as upper limit on the size.
}
\tablenotetext{e}
{
Estimates of the clump fractional contribution to the total
$H_{160}$ band emission of the galaxies.
$\mathcal{F}^{\rm cl}_{\rm PS}$ is the contribution assuming the clumps are
point sources, with the local background from the host galaxy subtracted.
The $\mathcal{F}^{\rm cl}_{\rm raw}$ fraction is derived from the flux in
the aperture of radius $r_{\rm phot}$, without subtraction of the host galaxy
background and without aperture correction.  These estimates represent lower
and upper limits, respectively, as argued in \S~\ref{Sub-clumps_meth}, and
are listed for comparison with the adopted background-subtracted,
aperture-corrected fractions simply denoted $\mathcal{F}^{\rm cl}$.
}
\tablenotetext{f}
{
Background- and aperture-corrected $H_{160}$ band magnitudes of the clumps
(uncorrected for extinction, in the AB system).
}
\tablenotetext{g}
{
Background- and aperture-corrected equivalent rest-frame $g$-band
luminosity of the clumps (uncorrected for extinction) relative to
the Sun's $g$-band luminosity density (which has an absolute
$M_{g,\odot} = 5.07~{\rm mag}$).
}
\tablenotetext{h}
{
Stellar mass estimates for the clumps, inferred from their light
contributions (background- and aperture-corrected) and the total
stellar mass derived from the SED modeling of each galaxy.
}
\end{deluxetable}
\tabletypesize{\small}

\clearpage

\tabletypesize{\scriptsize}
\begin{deluxetable}{ccccccccccccc}
\tablecolumns{13}
\tablewidth{0pt}
\tablecaption{$H_{160}$- and $i_{814}$-band magnitudes and color-based
              properties of clumps identified in the NIC2 and ACS images
              of MD\,41
              \label{tab-clumpsprop_md41}}
\tablehead{
   \colhead{Clump} &
   \colhead{$d_{\rm proj}$\,\tablenotemark{a}} &
   \colhead{$d$\,\tablenotemark{a}} &
   \colhead{$r_{\rm phot}$\,\tablenotemark{b}} &
   \multicolumn{4}{c}{From ``raw'' photometry\,\tablenotemark{c}} &
   \colhead{} &
   \multicolumn{4}{c}
    {From background-subtracted photometry\,\tablenotemark{d}} \\[0.9ex]
   \cline{5-8}  \cline{10-13} \\
   \colhead{} &
   \colhead{} &
   \colhead{} &
   \colhead{} &
   \colhead{$H_{160}^{\rm\,cl}$} &
   \colhead{$i_{814}^{\rm\,cl}$} &
   \colhead{$\log(M_{\star}/L_{g}^{\rm rest,\,cl})$} &
   \colhead{$M_{\star}^{\rm\,cl}$} &
   \colhead{} &
   \colhead{$H_{160}^{\rm\,cl}$} &
   \colhead{$i_{814}^{\rm\,cl}$} &
   \colhead{$\log(M_{\star}/L_{g}^{\rm rest,\,cl})$} &
   \colhead{$M_{\star}^{\rm\,cl}$} \\[0.5ex]
   \colhead{} &
   \colhead{(kpc)} &
   \colhead{(kpc)} &
   \colhead{(arcsec)} &
   \colhead{(mag)} &
   \colhead{(mag)} &
   \colhead{($\rm \log[M_{\odot}\,L_{\odot,\,g}^{-1}$])} &
   \colhead{($\rm 10^{8}\,M_{\odot}$)} &
   \colhead{} &
   \colhead{(mag)} &
   \colhead{(mag)} &
   \colhead{($\rm \log[M_{\odot}\,L_{\odot,\,g}^{-1}$])} &
   \colhead{($\rm 10^{8}\,M_{\odot}$)}
}
\startdata
  1  &  0.56   & $1.1 \pm 0.2$  &  0.10  &
  $26.20 \pm 0.04$  & $27.45 \pm 0.02$ &
   $-0.71^{+0.18}_{-0.17}$      &  $6.4^{+3.3}_{-2.1}$       & &
  $27.08 \pm 0.21$ & $> 29.65$        &
   $> -0.15$                    & $> 10.0$                   \\[0.5ex]
  2  &  1.65   & $2.3 \pm 0.2$  &  0.10  &
  $26.03 \pm 0.04$ & $27.16 \pm 0.02$ &
   $-0.75^{+0.17}_{-0.16}$      &  $6.8^{+3.2}_{-2.1}$       & &
  $26.82 \pm 0.17$ & $28.23 \pm 0.10$ &
   $\hphantom{< }-0.65^{+0.27}_{-0.23}$ & $\hphantom{< }4.2^{+3.6}_{-1.8}$ \\[0.5ex]
  3  &  2.96   & $3.2 \pm 0.1$  &  0.10  &
  $26.16 \pm 0.04$ & $27.03 \pm 0.02$ &
   $-0.89 \pm 0.17$             & $4.4^{+2.1}_{-1.4}$        & &
  $27.45 \pm 0.30$ & $28.84 \pm 0.17$ &
   $\hphantom{< }-0.65^{+0.35}_{-0.27}$ & $\hphantom{< }2.3^{+2.9}_{-1.3}$ \\[0.5ex]
  4  &  4.77   & $5.8 \pm 0.4$  &  0.10  &
  $26.26 \pm 0.05$ & $26.68 \pm 0.01$ &
   $-1.15 \pm 0.17$             & $2.2^{+1.1}_{-0.7}$        & &
  $26.59 \pm 0.13$ & $27.77 \pm 0.06$ &
   $\hphantom{< }-0.73^{+0.22}_{-0.20}$ & $\hphantom{< }4.3^{+2.9}_{-1.7}$ \\[0.5ex]
ACS-1 &  6.95   & $7.0 \pm 0.1$  &  0.10  &
  $27.16 \pm 0.11$ & $27.62 \pm 0.03$ &
   $-1.13^{+0.22}_{-0.20}$       &  $1.0^{+0.7}_{-0.4}$      & &
  $> 27.67$        & $28.64 \pm 0.14$ &
   $< -0.84$                     & $< 1.2$                   \\[0.5ex]
ACS-2 &  1.91   & $2.9 \pm 0.4$  &  0.10  &
  $26.73 \pm 0.07$ & $27.37 \pm 0.02$ &
   $-1.02 \pm 0.19$              & $1.9^{+1.1}_{-0.7}$       & &
  $> 27.67$        & $28.53 \pm 0.13$ &
   $< -0.90$                     &   $< 1.1$                 \\[0.5ex]
ACS-3 &  6.89   & $7.3 \pm 0.2$  &  0.15  &
  $26.46 \pm 0.09$ & $26.21 \pm 0.01$ &
   $-1.61 \pm 0.18$              &   $0.64^{+0.33}_{-0.22}$  & &
  $> 27.49$        & $26.78 \pm 0.03$ &
   $< -1.95$                     &   $< 0.11$                \\[-1.5ex]
\enddata
\tablecomments{
$H_{160}$ and $i_{814}$ photometry and color-based properties of clumps
in MD\,41, where the set of four clumps identified in the NIC2 image
(\#1 to \#4) is augmented with three clumps identified in the ACS map
but not in the NIC2 map (see Figure~\ref{fig-md41_panel}).
The ratio of stellar mass to rest-frame $g$-band luminosity (uncorrected
for extinction) of the clumps is computed from the relationship between
observed $i_{814} - H_{160}$ color and $M_{\star}/L_{g}^{\rm rest}$
derived in Paper~I (see also \S~\ref{Sub-case_md41}).
The clump stellar masses listed here are obtained from the color-based
$M_{\star}/L_{g}^{\rm rest}$ ratios inferred for individual clumps instead
of being based on a uniform ratio across the galaxy as implicitely assumed
for the estimates reported in Table~\ref{tab-clumpsprop}.
}
\tablenotetext{a}
{
Projected distance ($d_{\rm proj}$) relative to the geometric center
of the galaxies, and deprojected distance ($d$) assuming the clumps lie
in a disk of inclination with respect to the sky plane given by the
morphological axis ratio (Table~\ref{tab-targets}).
}
\tablenotetext{b}
{
Radius of the circular aperture used to measure the clump fluxes.
}
\tablenotetext{c}
{
Properties based on the fluxes measured in the circular aperture of
radius $r_{\rm phot}$, without subtraction of the local background from
the galaxy and without aperture correction.  The $1\,\sigma$ photometric
uncertainties are computed from the noise properties of the images.
The uncertainties for the $M_{\star}/L_{g}^{\rm rest}$ and $M_{\star}$
results account for those of the magnitudes and the accuracy of the
$i_{814} - H_{160}$ versus $M_{\star}/L_{g}^{\rm rest}$ relationship
($\rm \approx 0.15~dex$; see \S~\ref{Sub-case_md41_impl}).
}
\tablenotetext{d}
{
Properties based on the fluxes measured in the circular aperture of
radius $r_{\rm phot}$ with local background subtraction and aperture
correction.  For $r_{\rm phot} = 0\farcs 10$ and $0\farcs 15$, the
aperture correction is a factor of 2.09 and 1.63, respectively.
The $1\,\sigma$ photometric uncertainties reflect the noise properties
of the images and account for both the measurement of clump and background
fluxes; $3\,\sigma$ limits are given for clumps with background-subtracted
fluxes in one or the other band that are smaller than the $3\,\sigma$
photometric uncertainty.
The uncertainties for the $M_{\star}/L_{g}^{\rm rest}$ and $M_{\star}$
results are obtained in the same way as for those based on the raw
photometry.
}
\end{deluxetable}
\tabletypesize{\small}

\tabletypesize{\scriptsize}
\begin{deluxetable}{ccccccccccc}
\tablecolumns{11}
\tablewidth{0pt}
\tablecaption{$H_{160}$-band magnitudes, H$\alpha$ fluxes, and stellar ages
              of clumps identified in the NIC2 and SINFONI maps of BX\,482
              \label{tab-clumpsprop_bx482}}
\tablehead{
   \colhead{Clump} &
   \colhead{$d_{\rm proj}$\,\tablenotemark{a}} &
   \colhead{$d$\,\tablenotemark{a}} &
   \colhead{$r_{\rm phot}$\,\tablenotemark{b}} &
   \multicolumn{3}{c}{From ``raw'' photometry\,\tablenotemark{c}} &
   \colhead{} &
   \multicolumn{3}{c}
    {From background-subtracted photometry\,\tablenotemark{d}} \\[0.9ex]
   \cline{5-7}  \cline{9-11} \\
   \colhead{} &
   \colhead{} &
   \colhead{} &
   \colhead{} &
   \colhead{$H_{160}^{\rm\,cl}$} &
   \colhead{$F_{\rm H\alpha}^{\rm\,cl}$} &
   \colhead{Age} &
   \colhead{} &
   \colhead{$H_{160}^{\rm\,cl}$} &
   \colhead{$F_{\rm H\alpha}^{\rm\,cl}$} &
   \colhead{Age} \\[0.5ex]
   \colhead{} &
   \colhead{(kpc)} &
   \colhead{(kpc)} &
   \colhead{(arcsec)} &
   \colhead{(mag)} &
   \colhead{($\rm 10^{-17}~erg\,s^{-1}\,cm^{-2})$} &
   \colhead{(Myr)} &
   \colhead{} &
   \colhead{(mag)} &
   \colhead{($\rm 10^{-17}~erg\,s^{-1}\,cm^{-2})$} &
   \colhead{(Myr)}
}
\startdata
  1         &  4.74   & $4.8 \pm 0.2$  &  0.275  &
  $23.97 \pm 0.02$ & $3.92 \pm 0.05$ & $55 \pm 3$         & &
  $24.50 \pm 0.03$ & $2.41 \pm 0.05$ & $55^{+6}_{-5}$     \\[0.5ex]
  2         &  3.02   & $5.6 \pm 0.9$  &  0.15  &
  $26.02 \pm 0.05$ & $0.46 \pm 0.03$ & $115^{+26}_{-20}$  & &
  $27.30 \pm 0.30$ & $< 0.15$        & $> 94$             \\[0.5ex]
  3         &  4.36   & $6.7 \pm 1.3$  &  0.175  &
  $25.47 \pm 0.04$ & $1.14 \pm 0.03$ & $37^{+5}_{-4}$     & &
  $27.19 \pm 0.29$ & $0.17 \pm 0.03$ & $88^{+170}_{-47}$  \\[0.5ex]
  4         &  1.97   & $2.9 \pm 0.6$  &  0.20  &
  $25.14 \pm 0.03$ & $0.78 \pm 0.04$ & $239^{+40}_{-33}$  & &
  $26.48 \pm 0.16$ & $< 0.17$        & $> 570$            \\[0.5ex]
  5         &  3.68   & $7.6 \pm 1.0$  &  0.175  &
  $25.68 \pm 0.04$ & $1.09 \pm 0.03$ & $25^{+4}_{-3}$     & &
  $26.39 \pm 0.14$ & $0.43 \pm 0.03$ & $54^{+30}_{-17}$   \\[0.5ex]
H$\alpha$-1 &  3.69   & $5.8 \pm 1.1$  &  0.15  &
  $26.03 \pm 0.05$ & $0.57 \pm 0.03$ & $60^{+12}_{-10}$   & &
  $> 27.49$        & $0.25 \pm 0.03$ & $< 15$             \\[0.5ex]
H$\alpha$-2 &  6.31   & $6.6 \pm 0.8$  &  0.15  &
  $26.41 \pm 0.07$ & $0.42 \pm 0.03$ & $56^{+16}_{-12}$   & &
  $> 27.49$        & $0.19 \pm 0.03$ & $< 30$             \\[0.5ex]
\enddata
\tablecomments{
$H_{160}$-band and H$\alpha$ photometry and stellar age of clumps
in BX\,482, where the set of five clumps identified in the NIC2 image
(\#1 to \#5) is augmented with two clumps identified in the SINFONI H$\alpha$
map but not in the NIC2 map (see Figure~\ref{fig-bx482_panel}).
The stellar age is inferred from the rest-frame H$\alpha$ equivalent width
using \citet{BC03} evolutionary synthesis models for a constant star formation
rate, as described in \S~\ref{Sub-case_bx482}.
}
\tablenotetext{a}
{
Projected distance ($d_{\rm proj}$) relative to the geometric center
of the galaxies, and deprojected distance ($d$) assuming the clumps lie
in a disk of inclination with respect to the sky plane given by the
morphological axis ratio (Table~\ref{tab-targets}).
}
\tablenotetext{b}
{
Radius of the circular aperture used to measure the clump fluxes.
}
\tablenotetext{c}
{
Properties based on the fluxes measured in the circular aperture of
radius $r_{\rm phot}$, without subtraction of the local background from
the galaxy and without aperture correction.  The $1\,\sigma$ measurements
uncertainties are computed from the noise properties of the images.
The uncertainties for the ages account for those of the H$\alpha$
equivalent width measurements, propagated based on the CSF model
curves.
}
\tablenotetext{d}
{
Properties based on the fluxes measured in the circular aperture of
radius $r_{\rm phot}$ with local background subtraction and aperture
correction.  For $r_{\rm phot} = 0\farcs 15$, $0\farcs 175$, $0\farcs 20$,
and $0\farcs 275$, the aperture correction is a factor of 1.92, 1.68, 1.53,
and 1.28, respectively.
The $1\,\sigma$ photometric uncertainties reflect the noise properties
of the images and account for both the measurement of clump and background
fluxes; $3\,\sigma$ limits are given for clumps with background-subtracted
fluxes in one or the other map that are smaller than the $3\,\sigma$
photometric uncertainty.
The uncertainties for the ages are obtained in the same way as those
based on the raw photometry.
}
\end{deluxetable}
\tabletypesize{\small}

\tabletypesize{\small}
\begin{deluxetable}{llccccc}
\tablecolumns{7}
\tablewidth{0pt}
\tablecaption{Estimates of the contribution of a central stellar
              bulge component in the SINS NIC2 sample disks
              \label{tab-mass_core}}
\tablehead{
   \colhead{Property\,\tablenotemark{a}} &
   \colhead{Case\,\tablenotemark{b}} &
   \colhead{BX\,663} &
   \colhead{MD\,41} &
   \colhead{BX\,389} &
   \colhead{BX\,610} &
   \colhead{BX\,482}
}
\startdata
$\mathcal{F}^{\rm bulge}~(H_{160})$ & \ldots &
    0.017  &  0.035  &  0.034  &  0.012  &  0.013  \\[0.2ex]
$\mathcal{F}^{\rm bulge}~(M_{\star})$ &
    Passive $M/L$                           &  $0.018^{+0.000}_{-0.001}$  &
    $0.26^{+0.01}_{-0.02}$     &  $0.026 \pm 0.000$                       &
    $0.011 \pm 0.000$          &  $0.050 \pm 0.001$                  \\[0.2ex]
$\mathcal{F}^{\rm bulge}~(M_{\star})$ &
    Max. old SSP, $A_{V} = 0~{\rm mag}$     &  $0.039 \pm 0.001$          &
    $0.64^{+0.03}_{-0.04}$     &  $0.063 \pm 0.000$                       &
    $0.026 \pm 0.001$          &  $0.12 \pm 0.00$                    \\[0.2ex]
$\mathcal{F}^{\rm bulge}~(M_{\star})$ &
    Max. old SSP, $A_{V} = A_{V,{\rm SED}}$ &  $0.092^{+0.002}_{-0.003}$  &
    $2.34^{+0.10}_{-0.14}$     &  $0.19 \pm 0.00$                         &
    $0.063^{+0.001}_{-0.003}$   &  $0.28^{+0.00}_{-0.01}$            \\[0.2ex]
\enddata
\tablenotetext{a}
{
$\mathcal{F}^{\rm bulge}~(H_{160})$ is the estimated maximum contribution
to the total $H_{160}$ band emission of each galaxy of a putative hidden
bulge component with effective radius $R_{\rm e} = 1~{\rm kpc}$ and
S\'ersic index $n = 3$ (see \S~\ref{Sub-clumps_bulge}).
$\mathcal{F}^{\rm bulge}~(M_{\star})$ is the corresponding fraction of the
total galaxy stellar mass derived from the rest-UV/optical SED modeling.
Uncertainties on the latter reflect the ranges of fractional contributions
when varying the hidden bulge $R_{\rm e}$ up to 3~kpc, and $n$ between 2
and 4.  The range for the fraction of light for $n = 2$ and 4 are within
$< 0.5\%$ ot the values for $n = 3$ for all galaxies.  
}
\tablenotetext{b}
{
Different assumptions were made on the $M/L$ ratio of the hidden bulge:
a ratio corresponding to the median value of the massive quiescent
$z \sim 2$ galaxy sample of \citet{Dok08}, and that of a maximally-old
single SSP whose light is unobscured or attenuated by the same amount as
inferred from the SED modeling of each galaxy (see \S~\ref{Sub-clumps_bulge}).
}
\end{deluxetable}
\tabletypesize{\small}

\clearpage


\setcounter{figure}{0}

\begin{figure}[p]
\figurenum{1}
\epsscale{1.20}
\plotone{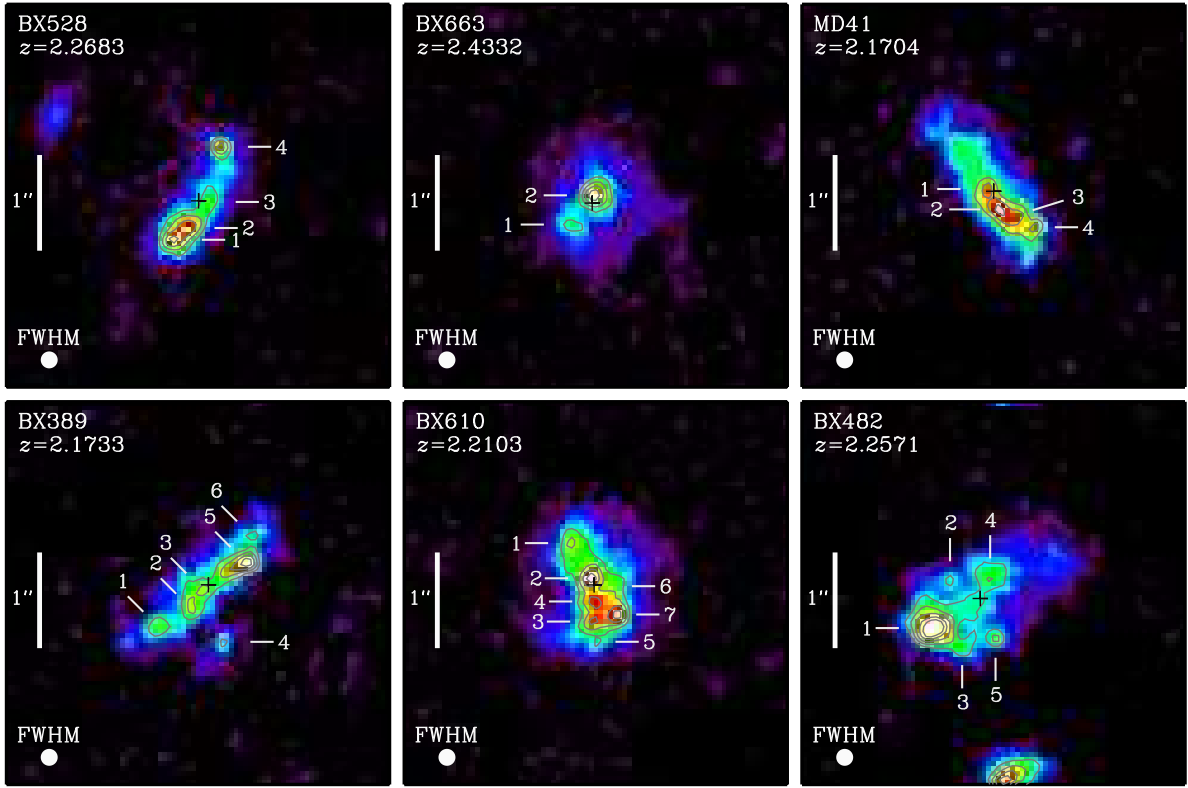}
\vspace{0.0cm}
\caption{
\small
{\em HST\/} NICMOS/NIC2 $H_{\rm 160}$-band maps of the galaxies, probing
the broad-band emission around rest-frame 5000\,\AA.  The target name
and H$\alpha$ redshift are labeled in the top corner of each panel.
The color coding scales linearly with flux density from black to white
for the minimum to maximum levels displayed (varying for each galaxy).
Contours are overplotted starting at flux densities $\approx 50\%$ of
the maximum in each image.
The FWHM of the effective PSF is indicated by the filled circle at the
bottom left of each corner; the angular resolution is $0\farcs 145$,
or $\rm \approx 1.2~kpc$ at the median $z = 2.2$ of the sources.
The angular scale of the images is shown with the 1\arcsec -long
vertical bar next to each galaxy.
In all maps, North is up and East is to the left.
The compact clumps identified in each galaxy are indicated,
with their ID number labeled.
\label{fig-clumpsid}
}
\end{figure}

\clearpage

\begin{figure}[!hpt]
\figurenum{2}
\epsscale{1.20}
\plotone{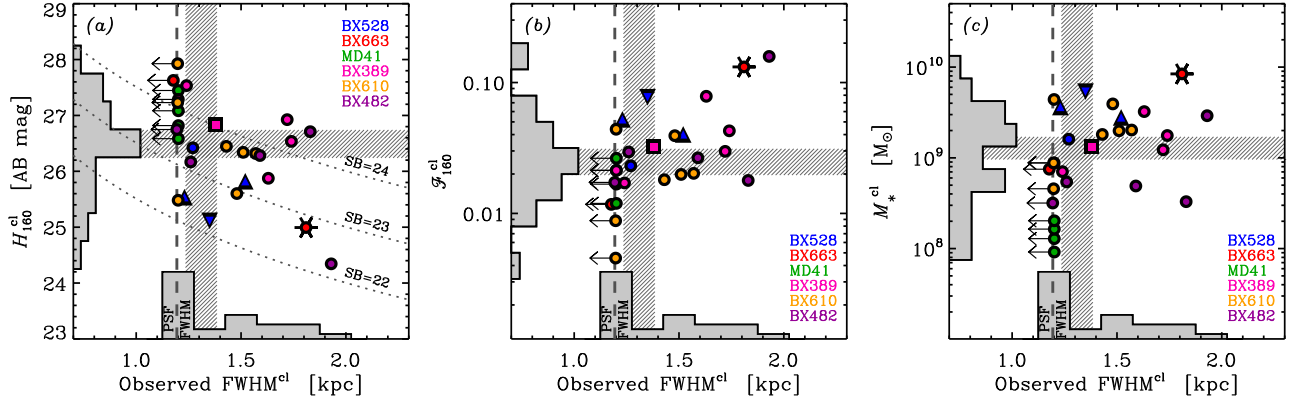}
\vspace{-0.5cm}
\caption{
\small
Properties of clumps identified in the NIC2 images of our galaxies.
{\em (a)\/} $H_{160}$ band magnitude versus observed FWHM.  
Dotted lines indicate curves of constant surface brightness as labeled,
in units of $\rm mag~arcsec^{-2}$.
{\em (b)\/} Fraction of the total galaxy $H_{160}$ band light measured
for each clump versus observed FWHM.
{\em (c)\/} Derived stellar mass of the clumps versus observed FWHM.
Different colors are used for clumps belonging to different galaxies,
following the scheme of the labels in each panel.  Clumps associated
with the southeast and northwest components of the merger system, BX\,528,
are plotted with triangles (upward and downward pointing, respectively),
the one corresponding to the southern companion of BX\,389 is plotted
with a square, and the one identified as the central peak of the AGN
galaxy BX\,663 is marked as starred circle.
In each panel, the histograms show the distributions of the measurements
projected onto the horizontal and vertical axes, and the hatched bars
correspond to the respective median values (including the upper limits
on the observed FWHMs).  The vertical dashed line shows the FWHM of the
PSF.
The observed FWHM of the clumps corresponds to the direct FWHM size;
for unresolved clumps, the PSF FWHM is taken as an upper limit on the
observed size.  Magnitudes and fractional light contributions are
measured in the circular apertures listed in Table~\ref{tab-clumpsprop},
with the local background from the host system subtracted and an
aperture correction based on the PSF profile applied, as described
in \S~\ref{Sub-clumps_meth}.
The stellar mass estimates are computed from the fractional light
contributions assuming a constant observed $M_{\star}/L_{\rm H160}$
ratio across each galaxy.
\label{fig-clumpsprop}
}
\end{figure}


\begin{figure}[!hpb]
\figurenum{3}
\epsscale{0.50}
\plotone{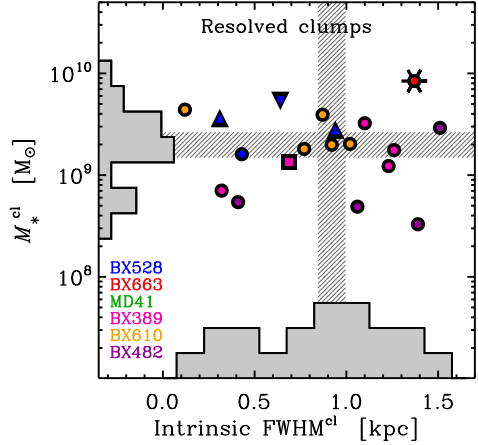}
\vspace{-0.0cm}
\caption{
\small
Stellar mass versus intrinsic FWHM size of clumps identified in
our NIC2 sample.  Data are plotted only for the resolved clumps,
i.e. with observed FWHM $>$ PSF FWHM.
Symbols are the same as used in Figure~\ref{fig-clumpsprop}, and
the histograms and hatched bars show the distributions and median
values for this set of resolved clumps.
\label{fig-clumpsprop2}
}
\end{figure}

\clearpage

\begin{figure}[!htp]
\figurenum{4}
\epsscale{1.10}
\plotone{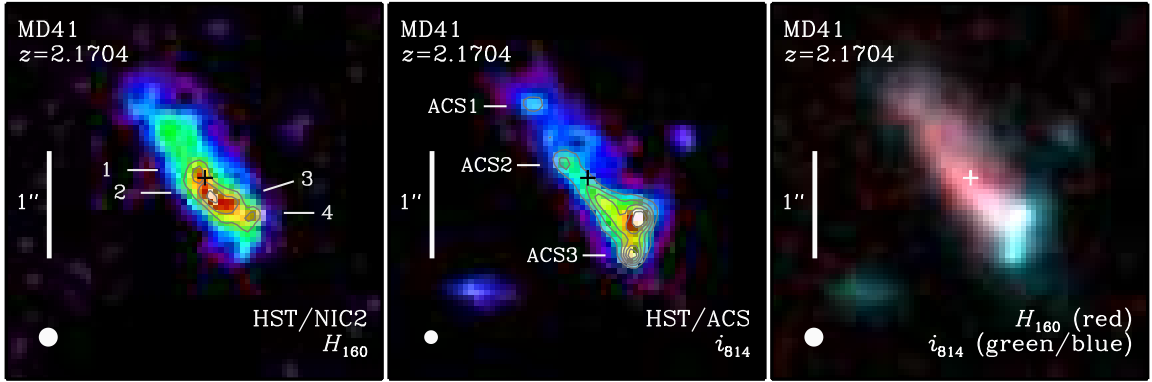}
\vspace{-1.0cm}
\caption{
\small
Comparison of rest-frame optical and UV morphologies for MD\,41.
North is up and east to the left in both images, and a bar of 1\arcsec\
in length indicates the scale in the images.
{\em (Left)\/} NIC2 $H_{160}$-band map, with the $\rm FWHM = 0\farcs 145$
of the PSF indicated by the filled circle at the bottom left corner.
{\em (Middle)\/} Same as the left panel but for the ACS $i_{814}$-band map,
at the original resolution of the data with PSF $\rm FWHM = 0\farcs 099$.
{\em (Right)\/} Color-composite after accurate PSF matching and registration
of the NIC2 and ACS maps; in this RGB image, the red channel is assigned to
the $H_{160}$ band, and the green and blue channels to the $i_{814}$ band.
The colors for all images plotted correspond to a linear flux density scale.
The white cross shows the location of the geometric center of MD\,41 as
determined from the NIC2 map.  Overall, the images are strikingly similar,
although on small scales there are noticeable differences.  In particular,
the two brightest clumps in the ACS image on the southwest edge of the
galaxy are much fainter in the NIC2 image and, consequently, have bluer
colors than the bulk of the source.
Clumps identified in the NIC2 $H_{160}$ band image are labeled in the
left panel, and those identified in the ACS $i_{814}$ band map that have
no counterpart in $H_{160}$ band emission are labeled in the middle panel.
\label{fig-md41_panel}
}
\end{figure}


\begin{figure}[!hbp]
\figurenum{5}
\epsscale{1.00}
\plotone{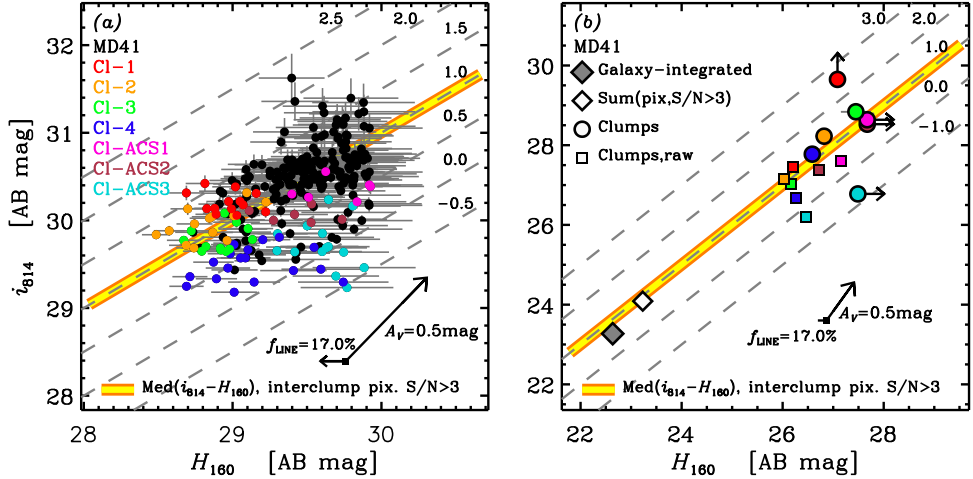}
\vspace{-0.7cm}
\caption{
\small
Observed near-IR $H_{160}$ band and optical $i_{814}$ band
magnitudes from the {\em HST\/} ACS and NIC2 imaging of MD\,41.
{\em (a)\/} Data for all individual pixels across MD\,41 with a
$\rm S/N > 3$ in both $H_{160}$ and $i_{814}$ bands.  Error bars
correspond to $1\,\sigma$ uncertainties on the measurements.
Pixels that fall within the photometric apertures of different clumps
are color-coded according to the scheme given by the labels.  Black
data points show pixels in the interclump regions.  Diagonal dashed
lines correspond to constant $i_{814} - H_{160}$ colors, as labeled
in the plot; the thick yellow-orange line shows the median color of
pixels in the interclump regions.  The arrows indicate the effects of
$A_{V} = 0.5~{\rm mag}$ of extinction on the magnitudes, and of an
emission line contribution in the $H_{160}$ bandpass of $17\%$ derived
from the integrated [\ion{O}{3}]\,$\lambda\lambda 4959,5007$ and H$\beta$
line fluxes for MD\,41 measured with SINFONI.
{\em (b)\/} Same as panel {\em (a)\/} but for the clump magnitudes.
Results based on the ``background-subtracted'' and ``raw'' photometry
(as explained in \S~\ref{Sub-case_md41_clumpsid}) are plotted as circles
and squares, respectively.  Error bars (mostly smaller than the symbol
size) correspond to the formal photometric uncertainties.  For clumps
undetected in one or the other band, $3\,\sigma$ upper limits on their
background-subtracted photometry are used.
For comparison, the integrated magnitudes of MD\,41 and those obtained
by summing up the fluxes of all individual pixels with $\rm S/N > 3$ are
plotted as grey- and white-filled diamonds, respectively.
The variations in colors across MD\,41 and among the clumps are fairly
modest for the most part; NIC2 clump \#1 and clump ACS-\#3 exhibit the
most significant differences.
The NIC2-identified clumps tend to be redder than those identified only
in the ACS image.
\label{fig-md41_pixplots}
}
\end{figure}

\clearpage

\begin{figure}[!htp]
\figurenum{6}
\epsscale{1.10}
\plotone{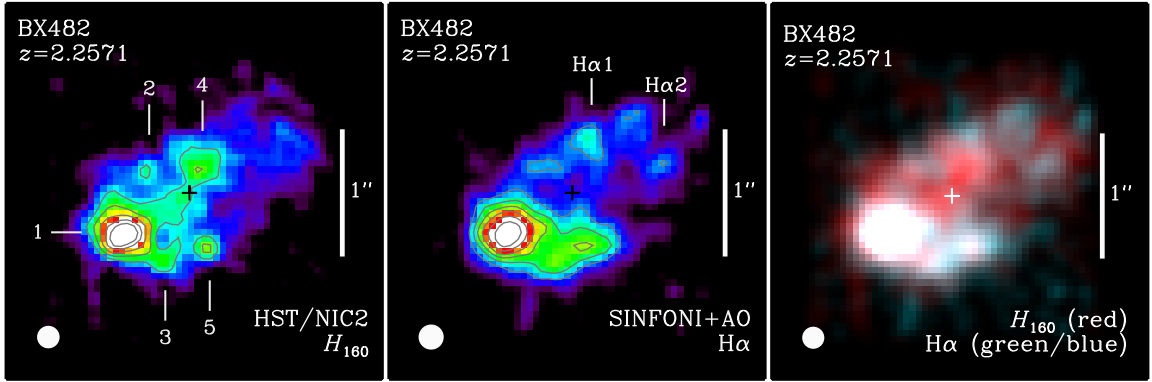}
\vspace{-1.0cm}
\caption{
\small
Comparison of H$\alpha$ and rest-frame optical morphologies for BX\,482.
North is up and east to the left in both images, and a bar of 1\arcsec\
in length indicates the scale in the images.
{\em (Left)\/} NIC2 $H_{160}$ band image at the original angular resolution
of the data, with the $\rm FWHM = 0\farcs 145$ of the PSF indicated by the
filled circle at the bottom left corner.
{\em (Middle)\/} Same as the left panel but for the SINFONI H$\alpha$ map,
with PSF $\rm FWHM = 0\farcs 17$.
{\em (Right)\/} Color-composite after accurate PSF matching and registration
of the NIC2 and H$\alpha$ maps; in this RGB image, the red channel is
assigned to the $H_{160}$ band and the green and blue channels to H$\alpha$.
The colors for all images plotted correspond to a linear flux density scale.
The white cross shows the location of the geometric center of BX\,482 as
determined from the NIC2 map.
The $H_{160}$ band and H$\alpha$ emission follow each other closely
although on small spatial scales there are some noticeable differences.
In particular, the clumpy structure on the northwestern side of the
galaxy center has less correspondence between the two maps than on the
southeastern side.
Clumps identified in the NIC2 $H_{160}$ band image are labeled in the
left panel, and those identified in the SINFONI H$\alpha$ line map that
have no counterpart in $H_{160}$ band emission are labeled in the middle
panel.
\label{fig-bx482_panel}
}
\end{figure}


\begin{figure}[!hbp]
\figurenum{7}
\epsscale{1.00}
\plotone{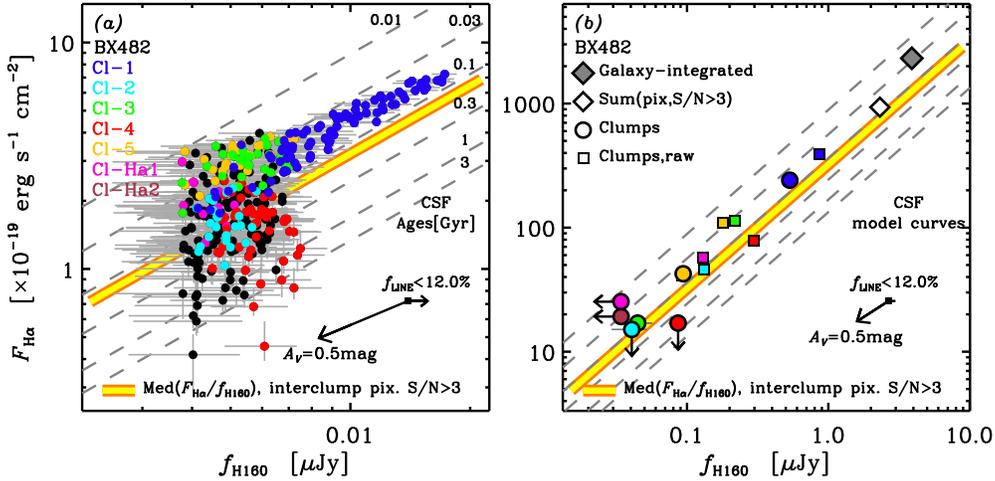}
\vspace{-0.7cm}
\caption{
\small
Observed near-IR $H_{160}$ flux density and H$\alpha$ line flux from
the {\em HST\/} NIC2 imaging and SINFONI$+$AO observations of BX\,482.
{\em (a)\/} Data for all individual pixels across BX\,482 with a
$\rm S/N > 3$ in both $H_{160}$ band and H$\alpha$ emission.
Error bars show the $1\,\sigma$ uncertainties on the measurements.
Pixels that fall within the photometric apertures of different clumps
are color-coded according to the scheme given by the labels, and black
data points show pixels in the interclump regions.  Diagonal dashed
lines correspond to constant rest-frame H$\alpha$ equivalent widths,
translated to stellar ages based on model predictions for constant
star formation (see \S~\ref{Sub-case_bx482_impl} for details); ages in Gyr
are labeled next to each line (and, from young to old ages, correspond
to $W^{\rm rest}({\rm H\alpha}) = 415$, 265, 170, 115, 80, and 60\,\AA).
The thick yellow-orange line shows the median age inferred for pixels
in the interclump regions.  The arrows indicate the effects of
$A_{V} = 0.5~{\rm mag}$ of extinction on the fluxes, and of the upper
limit for the contribution in the $H_{160}$ bandpass of $12\%$ derived
from the integrated fluxes of the [\ion{O}{3}]\,$\lambda\lambda 4959,5007$ 
emission lines and the $3\,\sigma$ limit for H$\beta$ measured in BX\,482
with SINFONI.
{\em (b)\/} Same as panel {\em (a)\/} but for the clump measurements.
Results based on the ``background-subtracted'' and ``raw'' photometry
(as explained in \S~\ref{Sub-case_bx482_clumpsid}) are plotted as circles
and squares, respectively.  Error bars (mostly smaller than the symbol
size) correspond to the formal photometric uncertainties.  For clumps
undetected in one or the other image, $3\,\sigma$ upper limits on their
background-subtracted photometry are used.
The integrated fluxes for BX\,482 and those summed up over all individual
pixels with $\rm S/N > 3$ are plotted as grey- and white-filled diamonds,
respectively.  The model curves are for the same ages as in panel {\em (a)\/}.
The clumps span a comparable range in
$W^{\rm rest}({\rm H\alpha})$, hence in inferred stellar ages, as the
interclump regions.  The NIC2-identified clumps tend to have lower ratios,
suggesting older ages, than those identified only in the H$\alpha$ map.
\label{fig-bx482_pixplots}
}
\end{figure}

\clearpage

\begin{figure}[p]
\figurenum{8}
\epsscale{0.60}
\plotone{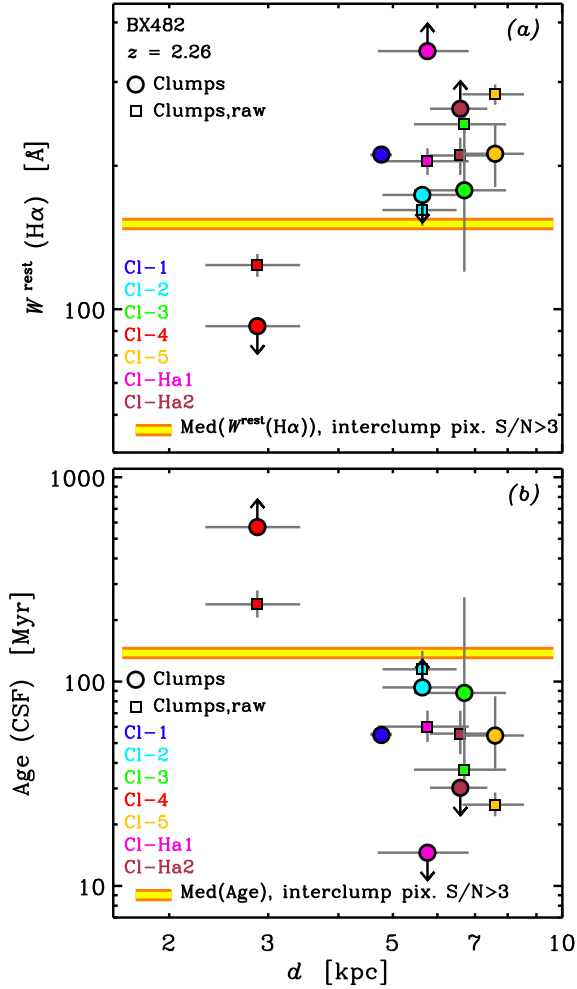}
\vspace{-0.0cm}
\caption{
\small
Properties as a function of deprojected galactocentric distance of clumps
in BX\,482.
{\em (a)\/} Rest-frame H$\alpha$ equivalent width, computed from the
SINFONI H$\alpha$ map and the NIC2 $H_{160}$ band image as described
in \S~\ref{Sub-case_bx482_ewha}.
Results based on the ``background-subtracted'' and ``raw'' photometry
are plotted as circles and squares, respectively.
Different colors indicate different clumps, as labeled in the plot.
Error bars correspond to the formal photometric uncertainties.  For
clumps undetected in one or the other image, $3\,\sigma$ upper limits
on their background-subtracted photometry are used.
The thick yellow-orange line shows the median $W^{\rm rest}({\rm H\alpha})$
measured for pixels in the interclump regions with a $\rm S/N > 3$ in each
map.
{\em (b)\/} Same as panel {\em (a)\/} but for the stellar ages derived
from comparison of the $W^{\rm rest}({\rm H\alpha})$ measurements to
model predictions for a constant star formation rate, as described in
\S~\ref{Sub-case_bx482_impl}.  The uncertainties of the ages account
for the photometric measurements uncertainties.
The clump closest to the center of BX\,482, NIC2-\#4, is clearly distinct
from the ensemble of the other clumps farther out in radius, with its
$W^{\rm rest}({\rm H\alpha})$ implying the oldest age of all clumps.
All other clumps have comparable or higher $W^{\rm rest}({\rm H\alpha})$,
hence younger inferred ages, than the median for the interclump regions.
For different star formation histories (SFHs), the absolute ages can change
significantly but the relative ages between clumps and interclump regions
are qualitatively unchanged as long as the different regions have similar
SFHs.
\label{fig-bx482_age_dproj}
}
\end{figure}

\clearpage

\begin{figure}[p]
\figurenum{9}
\epsscale{0.60}
\plotone{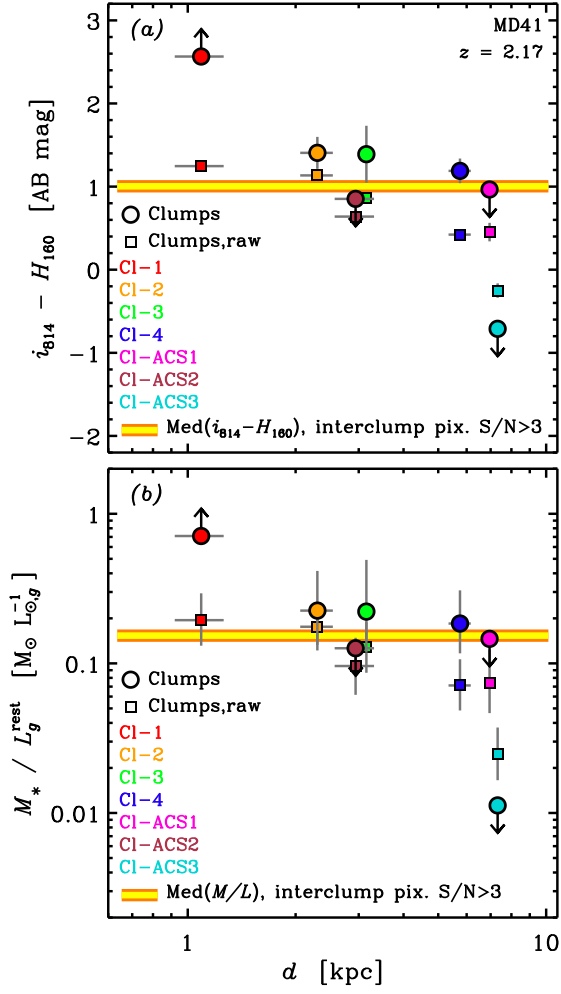}
\vspace{-0.0cm}
\caption{
\small
Properties as a function of deprojected galactocentric distance of clumps
in MD\,41.
{\em (a)\/} Observed $i_{814} - H_{160}$ colors, computed from the
ACS $i_{814}$ band and NIC2 $H_{160}$ band maps as described in
\S~\ref{Sub-case_md41_colvar}.
Results based on the ``background-subtracted'' and ``raw'' photometry
are plotted as circles and squares, respectively.
Different colors indicate different clumps, as labeled in the plot.
Error bars correspond to the formal photometric uncertainties.  For
clumps undetected in one or the other band, $3\,\sigma$ upper limits
on their background-subtracted photometry are used.
The thick yellow-orange line shows the median color measured for pixels
in the interclump regions with a $\rm S/N > 3$ in each band.
{\em (b)\/} Same as panel {\em (a)\/} but for the ratio of stellar mass
to dust-attenuated rest-frame $g$-band luminosity derived using the
relationship between observed $i_{814} - H_{160}$ colors and
$M_{\star}/L_{g}^{\rm rest}$ at $z = 2.2$, as described in
\S~\ref{Sub-case_bx482_impl} (see also \citeauthor*{FS11}).
The uncertainties of the $M/L$ ratios account for those of the
photometry as well as for the estimated accuracy of the color-$M/L$
relationship.
Clumps closer to the center of MD\,41 tend to have redder colors, hence
higher observed $M_{\star}/L_{g}^{\rm rest}$ ratios, which could reflect
increasing stellar ages and/or dust extinction of clumps at smaller radii.
\label{fig-md41_lmlg_dproj}
}
\end{figure}

\clearpage

\begin{figure}[p]
\figurenum{10}
\epsscale{1.0}
\plotone{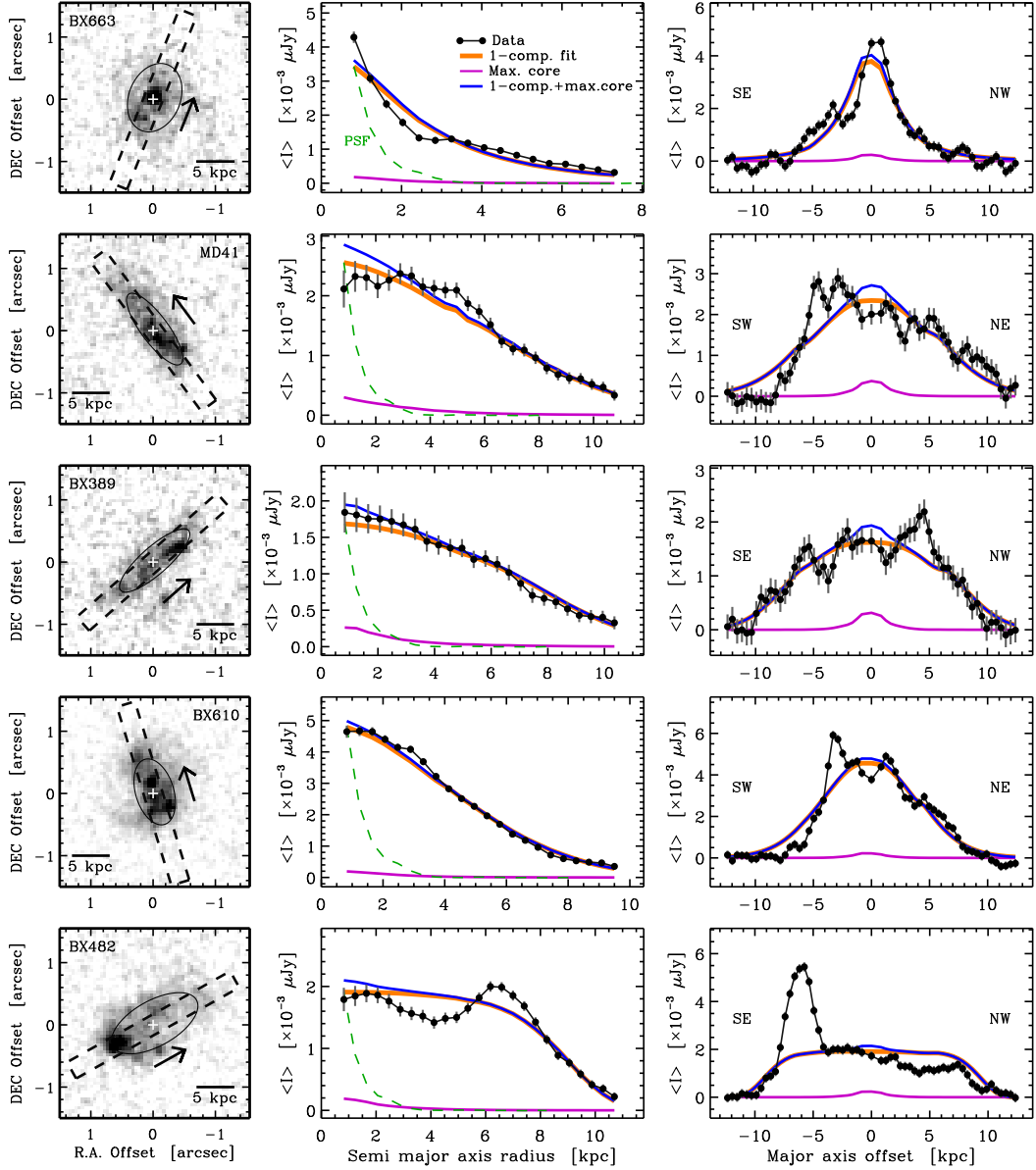}
\vspace{-0.5cm}
\caption{
\small
Images and light profiles extracted from the NIC2 images of the five
disks among our SINS NIC2 sample.  Each row corresponds to one galaxy,
and the panels are as follows.
{\em Left:} $H_{160}$ band maps on a linear greyscale intensity scale.
An ellipse is drawn, centered at the geometric center of each galaxy,
with semi-major axis length, axis ratio, and P.A. corresponding to the
best-fit parameters of the single-component S\'ersic model fits of
\citeauthor*{FS11}.  The rectangle shows the pseudo-slit used to extract
the major axis light profiles.  North is up, east is to the left.
{\em Middle:} Radial light profiles extracted in annuli of
increasing radius, with center, axis ratio, and P.A. given by the
best-fit single-component S\'ersic model (``disk'') to the $H_{160}$
band 2D surface brightness distribution of each galaxy.  The profiles
are shown for the data ({\em black solid line and filled dots\/}), the
best-fit ``disk'' component ({\em thick orange line\/}), the ``maximal''
bulge component with $R_{\rm e} = 1~{\rm kpc}$, $n = 3$, and $b/a = 1$
allowed by the $1\,\sigma$ measurements uncertainties at $r < 2~{\rm kpc}$
({\em magenta solid line\/}; see \S~\ref{Sub-clumps_bulge}), and the sum
of the ``disk'' and ``maximum bulge'' ({\em blue solid line\/}).
The circularly-symmetric light profile of the PSF is plotted for reference
({\em green dashed line\/}).
{\em Right:} Major axis light $H_{160}$ band light profiles extracted
extracted along the pseudo-slit of width $0\farcs 3$ shown in the left
panels; negative (positive) offsets relative to the center correspond
to the south (north) side of the galaxy.  Different colors for the
lines are the same as for the middle panels.
\label{fig-mass_core}
}
\end{figure}

\clearpage

\begin{figure}[p]
\figurenum{11}
\epsscale{1.0}
\plotone{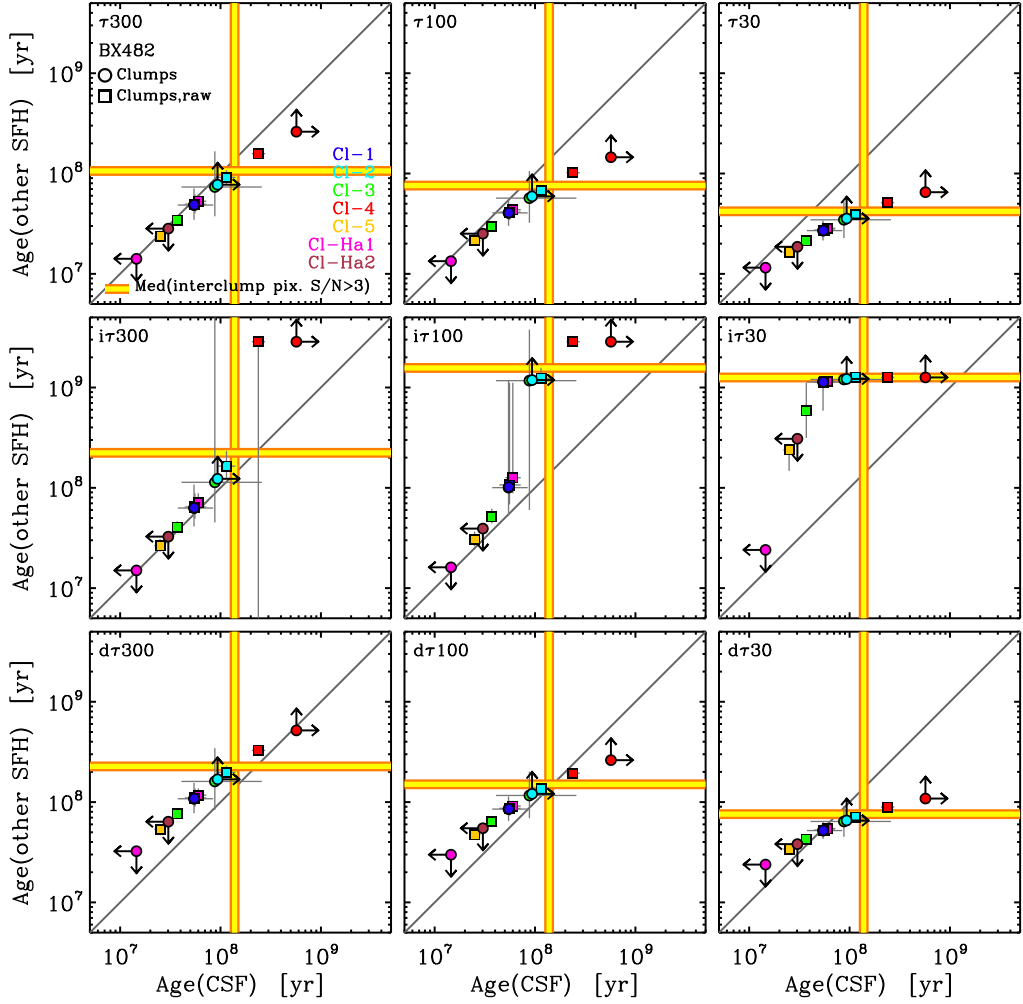}
\vspace{-0.5cm}
\caption{
\small
Comparison of stellar ages of clumps in BX\,482 derived from their
H$\alpha$ equivalent width for different assumptions on the star
formation history (SFH), as described in Appendix~\ref{App-WHa_models}.
In all panels, the horizontal axis corresponds to the age derived for
a constant star formation rate (CSF).  The vertical axis corresponds
to the age derived for alternative SFHs as follows.
{\em Top row\/}: exponentially-declining star formation rates
with $e$-folding timescales $\tau = 300$, 100, and 30~Myr
(labeled ``$\tau 300$,'' ``$\tau 100$,'' and ``$\tau 30$,''
respectively, in each panel).
{\em Middle row\/}: exponentially-increasing star formation
rates with the same $e$-folding timescales
(labeled ``i$\tau 300$,'' ``i$\tau 100$,'' and ``i$\tau 30$,''
respectively).
{\em Bottom row\/}: ``delayed'' star formation rates with the
same $e$-folding timescales
(labeled ``d$\tau 300$,'' ``d$\tau 100$,'' and ``d$\tau 30$,''
respectively).
Results based on the background-subtracted and raw clump photometry
are plotted as circles and squares, respectively.  Different colors
indicate different clumps, as labeled in the top left panel.
Uncertainties of the ages are computed by propagating the 
measurement uncertainties on the model curves.
The thick yellow-orange lines in each panel show the stellar age derived
from the median H$\alpha$ equivalent width of pixels in the interclump
regions with a $\rm S/N > 3$ in $H_{160}$ band and H$\alpha$ emission.
\label{fig-WHa_models}
}
\end{figure}

\end{document}